# Quantitative Nonlinear Optical Polarimetry with High Spatial Resolution

ALBERT SUCEAVA[1], SANKALPA HAZRA[1], JADUPATI NAG[1], JOHN HAYDEN[1], SAFDAR IMAM[2], ZHIWEN LIU[3], ABISHEK IYER[2], MERCOURI G. KANATZIDIS[2], SUSAN TROLIER-MCKINSTRY[1], JON-PAUL MARIA[1], VENKATRAMAN GOPALAN[1, 4, 5]*

[1] *Department of Materials Science and Engineering and the Materials Research Institute, The Pennsylvania State University, University Park, Pennsylvania 16802, USA*
[2] *Department of Chemistry, Northwestern University, Evanston, Illinois 60208, USA*
[3] *Department of Electrical Engineering, Pennsylvania State University, University Park, Pennsylvania, 16802, USA*
[4] *Department of Physics, Pennsylvania State University, University Park, Pennsylvania, 16802, USA*
[5] *Department of Engineering Science and Mechanics, Pennsylvania State University, University Park, Pennsylvania, 16802, USA*
*\*vxg8@psu.edu*

**Abstract:** Nonlinear optical microscopy such as in optical second harmonic generation (SHG) modality has become a popular tool today for probing materials in physical and biological sciences. While imaging and spectroscopy are widely used in the microscopy mode, nonlinear polarimetry that can shed light on materials' symmetry and microstructure is relatively underdeveloped. This is partly because quantitative analytical modeling of optical second harmonic generation (SHG) response for anisotropic crystals and films largely assumes low numerical aperture (NA) focusing of light where plane-wave approximation is sufficient. Tight focusing provides unique benefits in revealing out-of-plane polarization responses which cannot be detected by near plane-wave illumination in normal incidence. Here we outline a method for quantitatively analyzing SHG polarimetry measurements obtained under high-NA focusing within a microscope geometry. Experiments and simulations of a variety of standard samples, from single crystals to thin films are in good agreement, including measured and simulated spatial SHG maps of ferroelectric domains. A solution to the inverse problem is demonstrated, where the spatial distribution of a SHG tensor with unknown tensor coefficient magnitudes is determined by experimentally measured polarimetry. The ability to extract the out-of-plane component of the nonlinear polarization in normal incidence is demonstrated, which can be valuable for high resolution polarimetry of 2D materials, thin films, heterostructures and uniaxial crystals with a strong out-of-plane response.

## 1. Introduction

The detection of optical second harmonic generation (SHG) signal has become very popular in a wide swath of fields, from the identification of structural phases and phase transformations in crystalline materials, to the observation of biological molecular binding at interfaces [1–8]. SHG describes a process where two photons of the same frequency combine to generate a single photon upon interaction within an optically nonlinear material. It is a specific case of frequency-mixing process, the more general cases being sum frequency generation (SFG) and difference frequency generation (DFG) which occur when the interacting photons are at different frequencies [9]. The SHG process is expressed through a third-rank property tensor: $P_i^{2\omega} = \varepsilon_o d_{ijk} E_j^\omega E_k^\omega$, where $E_j^\omega$ and $E_k^\omega$ are the electric fields at frequency $\omega$ and polarizations $j$ and $k$ respectively, $P_i^{2\omega}$ is the nonlinear polarization generated in the material at frequency $2\omega$ and polarization, $i$, through the nonlinear tensor $d_{ijk}$ of the material, and $\varepsilon_o$ is the permittivity of vacuum. Neumann's principle determines the zero and non-zero elements of the SHG coefficients $d_{ijk}$, which reflect the symmetry of the material that this process occurs in [10]. As a direct consequence of this, the polarization-dependence of the SHG is highly sensitive to polar order and material symmetry.

The term SHG polarimetry refers to collecting various polarizations of the SHG signal as a function of the incident light polarization and is the most common kind of SHG experiment along with the Maker fringe method of measurement, where SHG intensity as a function of the incident probe angle is collected [11]. Detailed analysis of SHG polarimetry can allow for explicit determination of the local material structure and crystalline symmetry. Due to the high sensitivity of the SHG process to the relative orientations of crystalline axes and the polarization of the probe beam, more rigorous theoretical modeling is essential. Many works dedicated to this task exist, but primarily cover "tabletop" scale experiments where low numerical aperture (NA) lenses and a plane-wave approximation is used [11–15]. Such approximations are no longer valid, however, in a microscope setup where strongly focusing, high NA lenses are used. This work presents an examination of the problem of performing SHG polarimetry with high numerical aperture lenses, providing a theoretical model for analysis along with the measurement of a variety of well-studied reference materials, a comparison between experiment and theory, extraction of quantitative information about SHG coefficients for unknown materials, and the eventual generation of fully simulated SHG images alongside experimentally obtained images. Sensitive access to out-of-plane nonlinear polarization and the ability to extract spatially resolved maps of SHG tensors with unknown tensor coefficient magnitudes is demonstrated. The work could be a stepping stone for studies aimed at developing a more quantitative approach to analyzing SHG microscopy results in a wide range of physical and biological sciences.

## 2. SHG Microscopy

A scanning SHG microscope setup consists of a probe beam at the fundamental frequency ω focused onto a sample by a microscope objective, whereupon the SHG process occurs and light at the second harmonic frequency 2ω is collected by a highly sensitive detector such as a photomultiplier tube, as depicted in Fig. 1(a). For the purposes of performing polarimetry measurements, control over the polarization of the incident probe is required, typically achieved by using a half-waveplate, while an analyzer positioned in front of the detector allows for the selection of various polarized components of the SHG signal. The measurement is generally performed in normal incidence due to the constraints of using a high NA objective, but the experiment may be performed in a reflection or transmission geometry depending on the transmissivity of the sample and the positioning of the detector. In the case of the reflection geometry, a dichroic mirror is utilized to separate light at 2ω from the fundamental beam. Experiments discussed later in this work will occur within the reflection geometry at normal incidence. For further detailed discussion on the technique and a review of notable experiments on crystalline samples, readers are directed to Ref. [1].

It is important to differentiate between the multiple coordinate systems present in such an experiment: The orthogonal *lab coordinate system* refers to the arrangement and directions of the probe beam and optics involved, denoted using *X*, *Y*, and *Z* (Fig. 1(a)). The *crystal physics coordinate system*, 1, 2, and 3, represents the principal axes for the property tensor $d_{ijk}$ within the material. The *crystallographic coordinate system*, (*a*, *b*, *c*) represents three linearly independent crystallographic directions of the unit cell of a crystal that need not be orthogonal. These three coordinates in general need not be coincident [10,16]. The relationship between these three coordinate systems should be clearly established when performing experiments. When performing SHG polarimetry using a low NA, weakly focusing lens, it is sufficient to assume that the probe beam is a plane wave and takes on a single polarization state with the polarization perpendicular to the direction of propagation as depicted in Fig. 1(b). For lenses with more powerful focusing conditions however, this approximation no longer holds valid, and the polarization of the probe takes on a more complex character as different sections of the probe experience deflection in different directions, as depicted in Fig. 1(b-e).

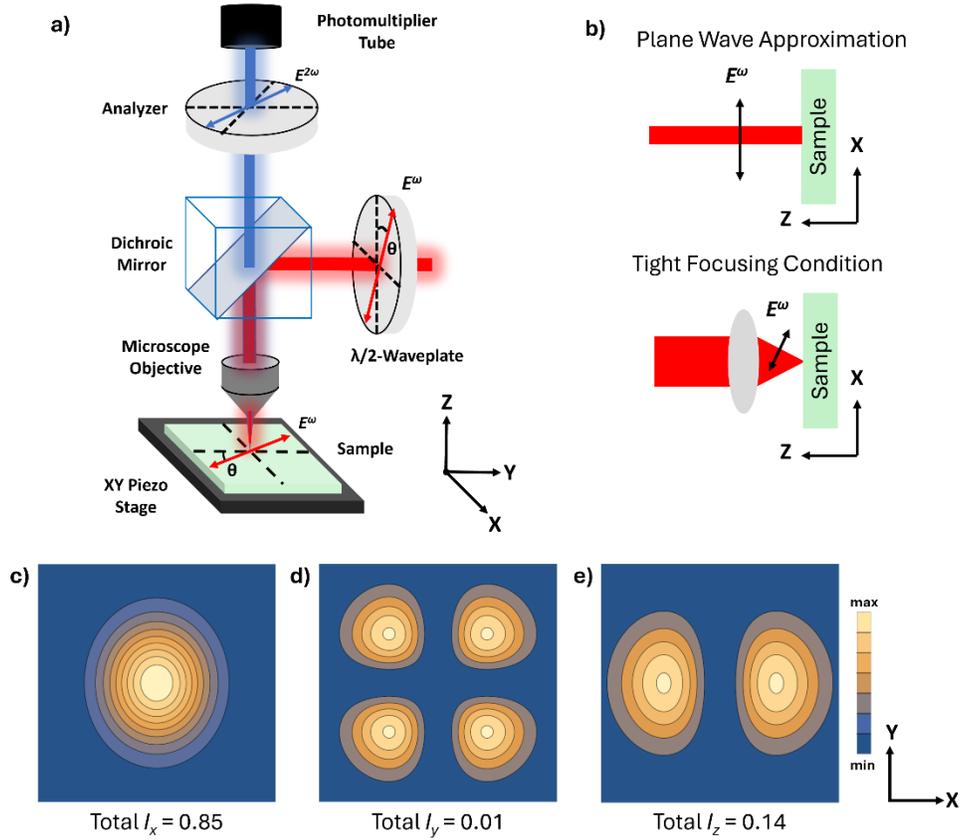

Fig. 1. a) Schematic representation of a typical SHG microscopy setup in a reflection configuration, with lab coordinate axes indicated. b) Schematic of the polarization of the probe beam when using low NA, high focal length lenses as opposed to a microscope objective. Contour plots of normalized intensity distributions for the $E_x$ (c), $E_y$ (d), and $E_z$ (e) components of a Gaussian probe beam linearly polarized along X within the focal plane of a NA = 0.75 objective obtained using the framework of section 3.

## 3. Theory

The problem of modeling SHG polarimetry as collected through a microscope involves solving for the incident and generated electric fields through sequential planes in the setup as presented in Fig. 2, with particular care taken towards the transformation of polarization vectors through the microscope objective. A vector model for polarized SHG microscopy is presented by Wang et al. and provides a strong basis for this work [17]. The integral forms for the optical electric field under a microscope as formulated by Richards and Wolf are employed by Wang et al. and are a common choice when beginning to model light-matter interactions in a microscope [18]. The work of Zhang et al. also presents an analytical form for vector field quantities of focused light in the context of SHG microscopy based on the Richards and Wolf approach, with additional consideration towards spherical aberration [19]. However, the approach of this work will instead follow the methodology of Leutenegger et al. which is based upon the plane wave spectrum method and Debye approximation to yield solutions that are flexible with regards to the input field profile, straightforward to implement numerically, and still converge to the Richards-Wolf representations [20].

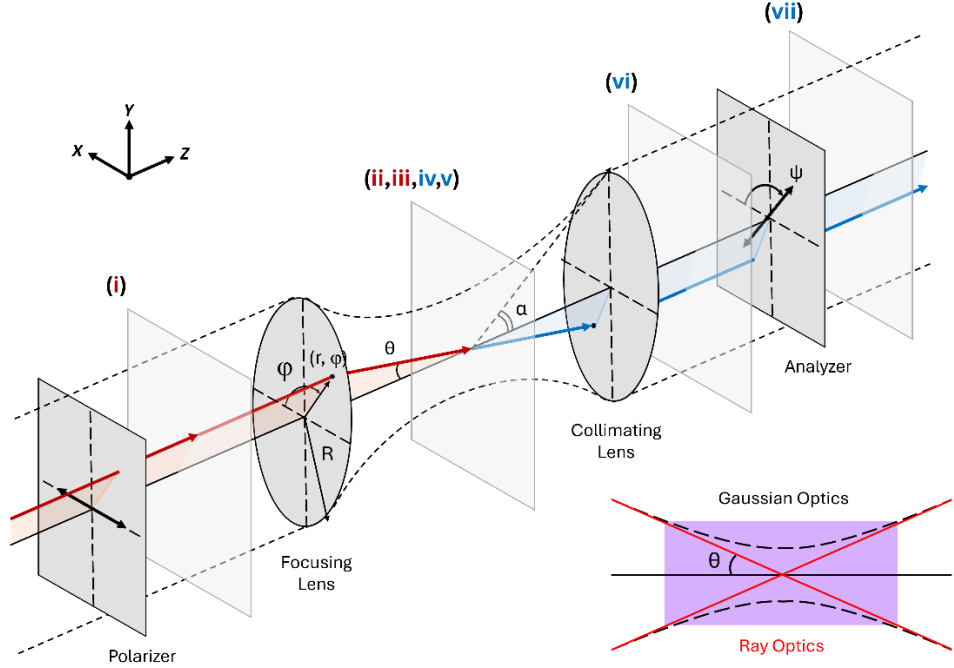

Fig. 2. Functional schematic of an SHG microscopy setup and the planes where optical electric fields are solved for. Red numerals indicate solutions for the fundamental fields while blue numerals indicate the second harmonic. The bottom right inset highlights the difference between a ray optics approach and true Gaussian beam behavior, with the purple region indicating the Rayleigh range where the two diverge. The approach of this work decomposes a Gaussian beam input profile into rays following plane-wave behavior, as depicted by the solid lines passing exactly through the focal point of the setup.

The form of the cross-sectional electric field profile of the probe beam $\vec{E}_{(i)}(r,\varphi)$, located at plane (i) in Fig. 2, is selected based on experiment; for example, the far-field profile of a Gaussian beam as will be the case for this work. For a beam interacting with a lens of aperture of radius R, the component of a beam at position $(r,\varphi)$ is deflected by an angle $\theta$ towards the focus, where $\sin\theta = \frac{r\,NA}{R\,n_\mu}$, $NA$ being the numerical aperture of the lens, $n_\mu$ the index of the surrounding medium, and R the radial dimension of the objective aperture. A uniform linear input polarization state is considered. One may consider the more generalized case of elliptically polarized light by decomposing it in the basis of two linearly polarized light and then proceeding as prescribed by this work. The readers are directed to Refs. [21–24] for understanding the advantages of SHG microscopy using circularly polarized light when imaging disordered or chiral samples. The incident field prior to interaction with the lens, $\vec{E}_{(i)}(r,\varphi)$, may be decomposed into radial and tangential components, which are the $p$ and $s$- polarized components $\hat{E}_p$ and $\hat{E}_s$ respectively upon focusing. Only the radial component experiences deflection and a change in polarization to become $\hat{E}_{p'}$:

$$\hat{E}_p = \begin{pmatrix} \cos\varphi \\ \sin\varphi \\ 0 \end{pmatrix} \tag{1a}$$

$$\hat{E}_s = \begin{pmatrix} -\sin\varphi \\ \cos\varphi \\ 0 \end{pmatrix} \tag{1b}$$

$$\hat{E}_{p'} = \begin{pmatrix} \cos\varphi \cos\theta \\ \sin\varphi \cos\theta \\ \sin\theta \end{pmatrix} \tag{1c}$$

which yields the transmitted plane wave spectrum immediately beyond the lens [20]:

$$\vec{E}_t(\theta,\varphi) = \left(\vec{E}_{(i)} \cdot \vec{E}_p\right)\hat{E}_{p'} + \left(\vec{E}_{(i)} \cdot \vec{E}_s\right)\hat{E}_s \tag{2}$$

An example of such fields obtained using a Gaussian input field profile, multiplied by their complex conjugate to yield intensities, is shown in Figure 1(c-e). The field at a point within the focal region, plane (ii), is obtained by performing a Debye diffraction integral of the transmitted waves over the maximum focusing angle of the lens, $\sin\alpha = NA/n_\mu$:

$$\vec{E}_{(ii)}(x,y,z) = -\frac{if}{\lambda_0}\int_0^\alpha \int_0^{2\pi} \vec{E}_t(\theta,\varphi) e^{i(k_z z - k_x x - k_y y)} \sin\theta \, d\varphi \, d\theta \tag{3}$$

where $f$ is the focal length of the lens and $\vec{k} = (k_x, k_y, k_z)$ is the wave vector in vacuum.

At this stage, the fields within the focal region given by $\vec{E}_{(ii)}(x,y,z)$ are sampled numerically at discrete points, either on a single plane or over an interaction volume, and used to populate a grid of values for each component of the electric field, $\vec{E}_j$ where $j = 1,2,3$ represent the crystal physics axes. We note $\vec{E}_{(ii)}(x,y,z) = \vec{E}_{(ii)}(r,\varphi,z)$ and the fields at every point $(r_{(ii)}, \varphi_{(ii)}, z_{(ii)})$ are computed by integrating the fields $\vec{E}_{(i)}$ as shown in Equation 3. The subscripts of $(r,\varphi,z)$ are henceforth suppressed for avoiding clutter and can be inferred from the subscripts of the fields with which they are associated.

This approach of going from $\vec{E}_{(i)}(r,\varphi)$ to $\vec{E}_{(ii)}(x,y,z)$ relies on deflecting parts of the incident probe profile towards the origin based on their distance from the optical axis of the lens normalized against the aperture. One can attempt to model experimental nonidealities such as underfilling and decentering of the beam, by tuning the input beam profile $\vec{E}_{(i)}(r,\varphi)$ accordingly. While the approach of Ref. [20] assumes an aplanatic system, for which decentering and tilt will render invalid, the accuracy with which this approach can approximate field profiles under such effects is discussed in the Supplemental information and Fig. S1.

With the fields at the focal plane sampled numerically, one can view each point as a ray with polarization given by the components $\vec{E}_j$ and the amplitude of the ray field reflected in the non-normalized values of these polarization components. It remains to evaluate the fundamental SHG equation for each ray before integrating the emitted second harmonic fields over the collecting microscope objective. To account for interaction with a single material interface, the individual components of the electric field of each ray are weighted by the Fresnel coefficients for $p$ and $s$-polarized light appropriately. As each ray is associated with a wavevector defining a unique plane of incidence, this is accomplished by identifying a rotation about the Z-axis, described by the rotation matrix $M$, such that each wavevector lies entirely in the XZ-plane, and then applying the same rotation to the associated electric field vector. This rotated electric field vector, $\vec{E}'_{(ii)} = M \cdot \vec{E}_{(ii)}$, is then weighted by the Fresnel coefficients:

$$M = \begin{pmatrix} \cos\varphi & \sin\varphi & 0 \\ -\sin\varphi & \cos\varphi & 0 \\ 0 & 0 & 1 \end{pmatrix} \quad (4)$$

$$\vec{E}'_{(iii)} = \begin{pmatrix} t_p(\theta)E'_{(ii),x} \\ t_s(\theta)E'_{(ii),y} \\ t_p(\theta)E'_{(ii),z} \end{pmatrix} \quad (5)$$

before being subjected to the operation by the inverse rotation matrix $\vec{E}_{(iii)} = M^{-1} \cdot \vec{E}'_{(iii)}$. The incorporation of the Fresnel coefficients is the sole instance where the sample refractive index is considered and thus of crucial importance. The variable $\varphi$ in Equation (4) refers to azimuthal coordinate on plane (ii), i.e. $\vec{E}_{(ii)}(r_{(ii)}, \varphi_{(ii)}, z_{(ii)})$. The variable $\theta$ in Equation (5) refers to the incident angle of the individual focused ray in plane (ii) shown in Figure 2. There is a one-to-one correspondence in the sampling space between plane (ii) immediately before the sample surface and plane (iii) immediately after the sample surface. i.e. $(r_{(ii)}, \varphi_{(ii)}, z_{(ii)}) \rightarrow (r_{(iii)}, \varphi_{(iii)}, z_{(iii)})$.

The methodology of this work returns to the approach outlined by Wang et al. and Török et al. [17,25]. The induced second harmonic signal follows from evaluating $P_i^{2\omega} = \varepsilon_0 d_{ijk} E_j^\omega E_k^\omega$ for each $\vec{E}_{(iii)}$ associated with a discrete ray sampled in the previous step.

This assumes that the form of the property tensor $d_{ijk}$ is known beforehand. For crystalline samples, the material structure and therefore symmetry can be obtained most directly from complementary X-ray diffraction measurements. For amorphous or polycrystalline samples that still possess texture and anisotropy, the material symmetry can be described using Curie groups, some of which possess third-rank property tensors with nonzero elements [10]. For SHG from centrosymmetric materials, where the dominant anisotropic response originates from surface symmetry breaking, one needs only use the form of the property tensor describing the surface symmetry. As opposed to a three-dimensional space group, a two-dimensional lattice may be described by one of 80 layer groups with each layer group possessing a particular point group that identifies the form of the SHG property tensor for that surface [26]. The layer group of the surface may be recognized from the bulk point group by identifying the symmetry elements with projections parallel to the surface that may be broken. This assumes the surface is smooth and uniform, as opposed to a nanostructured surface with complex morphology. Surface reconstruction, as often occurs, might further change the symmetry of the surface [27]. Other contributions from such materials, like the bulk quadrupolar response, is an example of a higher order nonlinear interaction beyond the immediate scope of this work but discussed briefly in the Supplementary material.

The emitted second harmonic radiation is then described by:

$$\vec{E}_{(iv)} = \frac{\exp(2ik|\vec{r} - \vec{r}'|)}{4\pi|\vec{r} - \vec{r}'|} \left\{ \hat{k} \times \left[ \hat{k} \times \overrightarrow{P^{2\omega}}(\vec{r}') \right] \right\} \quad (6)$$

where $\overrightarrow{P^{2\omega}}(\vec{r}')$ refers to induced nonlinear dipoles at $\vec{r}'$ for each discrete ray sampled and $\hat{k}$ in Equation (6) is the refracted wavevector beyond the sample interface [17]. This form of the radiated fields from a point nonlinear dipole source is consistent with what is obtained from a Green's function-based formalism [28–32]. A more detailed derivation is presented in the Supplement. Again, the effect of a single interface is considered in the same fashion as the evolution from $\vec{E}_{(ii)}$ to $\vec{E}_{(iii)}$. The rotation $M$ is used to bring $\vec{k}$ in the XZ-plane and then the same rotation applied to $\vec{E}_{(iv)}$: $\vec{E}'_{(iv)} = M \cdot \vec{E}_{(iv)}$. Then, $\vec{E}'_{(v)}$ is obtained by applying the Fresnel coefficients for a wave exiting the material, analogous to the operations on $\vec{E}'_{(ii)}$ to yield $\vec{E}'_{(iii)}$, before the inverse rotation $\vec{E}_{(v)} = M^{-1} \cdot \vec{E}'_{(v)}$ is applied.

$$\vec{E}'_{(v)} = \begin{pmatrix} t_p(\theta) E'_{(iv),x} \\ t_s(\theta) E'_{(iv),y} \\ t_p(\theta) E'_{(iv),z} \end{pmatrix} \tag{7}$$

The resultant discrete SHG dipole sources for every ray are summed over as if integrating, yielding a point source for the emitted SHG [17,25].

The polarization of the generated second harmonic light will experience a deflection again as it passes through the collimating lens, which in a reflection geometry is the same microscope objective that focuses the probe beam. The deflection of the beam follows [25]:

$$\vec{E}_{(vi)} = R^{-1} \cdot L^{-1} \cdot R \cdot \vec{E}_{(v)} \tag{8}$$

where $R$ and $L$ are rotation matrices about the optical axis of the system (defined as the lab $Z$ direction in Fig. 2) and about the direction the polarization is deflected against (the lab $Y$ direction for a ray rotated by $R$ to lie along the lab $X$) respectively [25]:

$$R = \begin{pmatrix} \cos\varphi & \sin\varphi & 0 \\ -\sin\varphi & \cos\varphi & 0 \\ 0 & 0 & 1 \end{pmatrix} \tag{9a}$$

$$L = \begin{pmatrix} \cos\theta & 0 & \sin\theta \\ 0 & 1 & 0 \\ -\sin\theta & 0 & \cos\theta \end{pmatrix} \tag{9b}$$

such that

$$\begin{aligned} E_{(vi),x} &= -E_{(v),x} \sin^2\varphi - \cos\theta\cos\varphi \left( E_{(v),x}\cos\varphi + E_{(v),y}\sin\varphi \right) \\ &\quad + \cos\varphi \left( E_{(v),z}\sin\theta + E_{(v),y}\sin\varphi \right) \end{aligned} \tag{10a}$$

$$\begin{aligned} E_{(vi),y} &= E_{(v),z}\sin\varphi\sin\theta + \cos\varphi \left( E_{(v),y}\cos\varphi + E_{(v),x}\sin\varphi \right) \\ &\quad - \cos\theta\sin\varphi \left( E_{(v),x}\cos\varphi + E_{(v),y}\sin\varphi \right) \end{aligned} \tag{10b}$$

$$E_{(vi),z} = 0 \tag{10c}$$

An analyzer placed in front of the detector can select for a specific polarized component of the generated SHG. If the polarizer is placed in at an arbitrary angle $\psi$ [17]:

$$E_{(vii),x} = \cos^2\psi\, E_{(iv),x} + \sin\psi\cos\psi\, E_{(vi),y} \tag{11a}$$
$$E_{(vii),y} = \sin\psi\cos\psi\, E_{(iv),x} + \sin^2\psi\, E_{(vi),y} \tag{11b}$$
$$E_{(vii),z} = 0 \tag{11c}$$

The final signal observed by the detector is obtained by integrating the second harmonic fields over the collection angle of the collimating lens, $\alpha$ [17]:

$$I^{2\omega} = \int_0^{2\pi} d\varphi \int_0^{\alpha} d\theta \left| \vec{E}_{(vii)} \right|^2 r^2 \sin\theta \tag{12}$$

It is important to note that this approach is based upon geometrical ray optics and does not account for the full theoretical behavior of a focused Gaussian beam within the Rayleigh range. To summarize, a chosen input beam profile is decomposed into many rays, each a packet of information containing vector quantities for the wavevector and electric field. Each ray is deflected by the lens, transforming the vectors, and phase accumulated via the diffraction integral. These rays are used to generate second harmonic fields which experience analogous transformations before collection by a detector. This approach would imply that all rays focus to a single point, at odds with the concept of a finite Gaussian beam waist and the plane wave-like behavior expected by a Gaussian beam at the exact focus. It will be noted in the following sections that tight focusing of the fundamental beam can lead to SHG polarimetry which can be explained by the out-of-plane component of the focused fundamental electric field, something that would not be expected for a thin sample at the focus of a Gaussian beam. The ray optics approach is expected to converge with true Gaussian beam behavior for those portions of the beam immediately out of the Rayleigh range, where the divergent rays closely follow the focused beam profile. Based on agreement between modelling and experimental results, these regions of the beam can be imagined as still contributing significantly towards generating the observed second harmonic signal.

It is relevant to emphasize that this framework seeks to model the SHG response from a single interface accessed in a reflection geometry. As a result, the sample interaction volume is considered to be small and the effects of temporal and spatial walk off are neglected [11]. Imaging performed in a transmission mode geometry would be affected by beam walk-off, phase matching, and multi-reflection effects which are beyond the scope of the current work to model under the tight focusing condition. Readers are invited to consult Ref. [15] for a plane-wave approximation-based framework that does account for these effects. Further justification for performing SHG microscopy only when focused on a sample interface is provided in Supplemental Figure S2 and the accompanying text. While the following discussion will discuss the SHG response from thin film and bulk crystalline materials, extensions to other nonlinear processes can be obtained by changing the solution of $\vec{E}_{(iv)}$. This is covered in further detail in the Supplementary Material.

## 4. Results and discussion

### 4.1 Experimental Polarimetry on Single Crystals Using High NA Objectives

To illustrate the necessity for such an analytical approach, polarimetry experiments are carried out on several reference samples on both a standard tabletop setup (0.12 NA) at normal incidence as well as under a microscope (0.75 NA). SHG microscopy experiments are performed using a Spectra-Physics Solstice Ace Ti:Sapphire seed laser, with fundamental wavelength at 800 nm, a repetition rate of 80 MHz, and pulse duration of 80 fs, chopped mechanically at 1.2 kHz for lock-in detection. For the tabletop case, the amplified laser of the same system is used, resulting in a repetition rate of 1 kHz and a pulse duration of 100 fs at the same wavelength. A uniform linear input polarization is achieved for the probe by using a Glan-Laser prism with a 100,000:1 extinction ratio (Thorlabs GL10). A similar prism is placed before a PMT (Hamamatsu H7826) to select the polarization state of the detected SHG light. It is worth stressing that poor control of polarization, either for the fundamental probe or in SHG detection, is one of the most common reasons for experimental data to diverge from theoretical predictions. All samples measured in the following sections are thin films deposited on polished substrates or single crystals procured with optical grade polished surfaces to avoid potential depolarization effects arising from scattering off rough interfaces. Such effects can result in the destruction of information regarding the sample anisotropy by disrupting the polarization of the probe or detected SHG, and thus are essential to eliminate when performing SHG polarimetry.

***z-cut α-quartz***: The first reference sample selected is *z*-cut α-quartz purchased from MTI Corporation, where *z*- refers to the crystallographic axis, [0001], oriented out-of-plane (the *z*-, used by industry, should not be confused with the lab coordinate *Z* in Fig. 1). The crystal possesses the point group 32 with SHG tensor:

$$\text{Point group 32: } d_{ijk} = \begin{pmatrix} d_{11} & -d_{11} & 0 & d_{14} & 0 & 0 \\ 0 & 0 & 0 & 0 & -d_{14} & -d_{11} \\ 0 & 0 & 0 & 0 & 0 & 0 \end{pmatrix} \tag{13}$$

where the tensor is expressed in the abbreviated Voigt notation. This notation relies upon interchangeability of the electric field terms in the SHG equation: $P_i^{2\omega} = \varepsilon_o d_{ijk} E_j^{\omega} E_k^{\omega}$. Since $E_j^{\omega} E_k^{\omega} = E_k^{\omega} E_j^{\omega}$, $d_{ijk} = d_{ikj}$ follows and thus one can contract the *j* and *k* subscripts as: $11 = 1, 22 = 2, 33 = 3, 23 = 32 = 4, 13 = 31 = 5$, and $12 = 21 = 6$ to account for this inherent symmetry [1].

The greatest source of differences in polarimetry measurements between the two experiments lies in the ability to sample the out-of-plane component of the nonlinear polarization when using a high NA lens. For a material of point group 32 however, when measuring SHG at normal incidence along the optic axis there exists no out-of-plane polarization to sample; $d_{11}$ describes the generation of a nonlinear polarization along the crystal physics 1 direction in response to fields also along 1, while $d_{14}$ describes a polarization along the 1 direction in response to fields along the 2 and 3 directions. In quartz specifically, $d_{11}$ = 0.3 pm/V and $d_{14}$ = 0.008 pm/V, leading to $d_{14}$ often being neglected entirely: $d_{14} \sim 0$ [9]. That is to say that measurement under a microscope as opposed to the tabletop should not yield any difference in polarimetry for this sample and any observed differences may instead be attributed to the experimental setup itself. The results of measurement in the two cases are shown in Fig. 3(a,b). It is apparent that the two curves are nearly identical, save for a slight inflation of the minima in the case of the microscope measurement.

The chosen geometry of measuring *z*-cut quartz in a normal incidence, transmission configuration with plane wave approximation results in the following analytical forms of polarimetry curves:

$$I_X^{2\omega} \propto |P_X^{2\omega}|^2 \propto d_{11}^{\ 2} t_{\omega}^4 \cos^2 2\varphi \tag{14a}$$
$$I_Y^{2\omega} \propto |P_Y^{2\omega}|^2 \propto d_{11}^{\ 2} t_{\omega}^4 \sin^2 2\varphi \tag{14b}$$

for the case of the [11$\bar{2}$0] crystallographic direction aligned with the lab *X*-axis, where $t_{\omega}$ is the Fresnel transmission coefficient at normal incidence. The maxima observed reflects the magnitude of the $d_{11}$ coefficient, with natural minima reaching zero, reflecting a geometry where generated SHG is polarized orthogonal to the screening analyzer. The plane wave fits are compared to experiments in Figs. 3(a) and (b), showing that they are good fits for low NA polarimetry data, while they are reasonable for the high NA polarimetry primarily because the out-of-plane source component is small in this case.

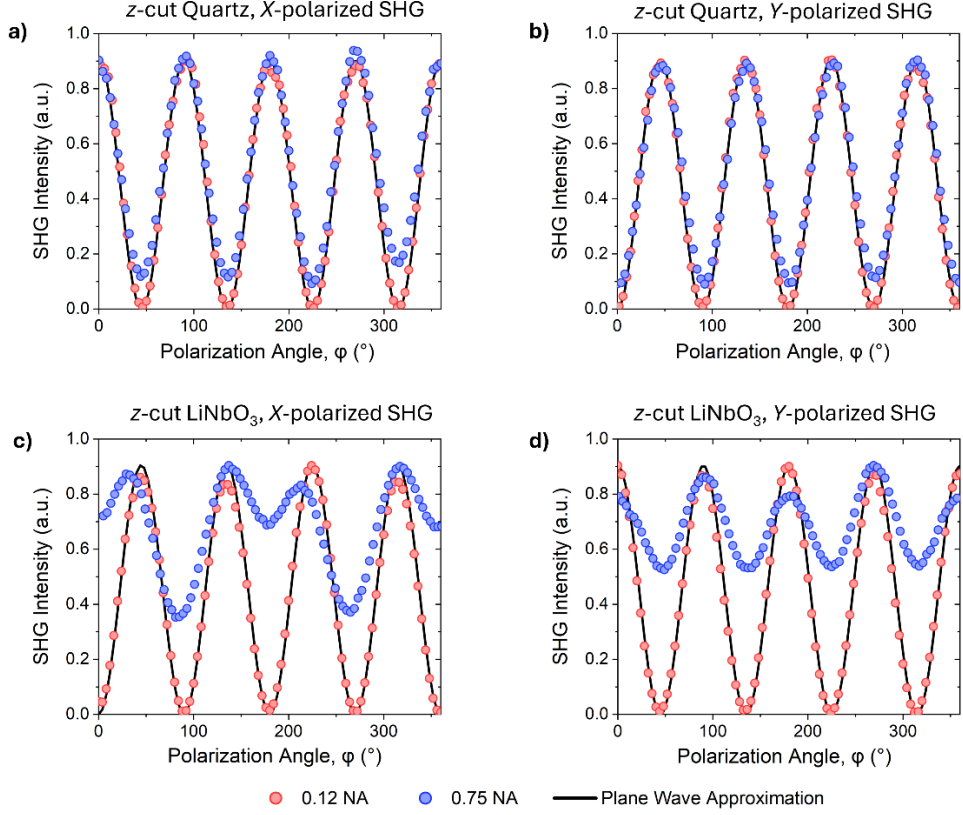

Fig. 3. SHG polarimetry curves obtained for single crystal *z*-cut quartz (a,b) and *z*-cut LiNbO$_3$ (c,d) with a 0.12 NA lens (red) and under a microscope with a 0.75 NA objective (blue). Curves on the left are for the analyzer oriented along the *X*-axis of the setup, aligned with the [11$\bar{2}$0] crystal direction while curves on the right are for the analyzer oriented along the *Y*-axis, aligned with the crystallographic [10$\bar{1}$0].

*z-cut LiNbO$_3$*: The second sample referenced is *z*-cut LiNbO$_3$, (where the *z*- again refers to the [0001] direction by industry standard; *z*- should not be confused with *Z* axis in Fig. 1) also purchased from MTI Corporation, representing a well characterized nonlinear optical material of point group 3*m*, with $d_{33}$ = -27.2 pm/V, $d_{31} = d_{15}$ = -4.35 pm/V, and $d_{22}$ = 2.1 pm/V [33].

$$\text{Point group } 3m: d_{ijk} = \begin{pmatrix} 0 & 0 & 0 & 0 & d_{15} & -d_{22} \\ -d_{22} & d_{22} & 0 & d_{15} & 0 & 0 \\ d_{31} & d_{31} & d_{33} & 0 & 0 & 0 \end{pmatrix} \quad (15)$$

With the probe polarization entirely in plane, the observed SHG signal on the tabletop will originate entirely from $d_{22}$, analogous to measurement of $d_{11}$ from *z*-cut quartz:

$$I_X^{2\omega} \propto |P_X^{2\omega}|^2 \propto d_{22}{}^2 t_\omega^4 \sin^2 2\varphi \quad (16a)$$
$$I_Y^{2\omega} \propto |P_Y^{2\omega}|^2 \propto d_{22}{}^2 t_\omega^4 \cos^2 2\varphi \quad (16b)$$

Unlike quartz however, LiNbO$_3$ possesses a dominant $d_{33}$ coefficient that cannot be sampled when probed along the optic axis with a low NA objective but will greatly influence the shape of a polarimetry curve collected under a microscope with a high NA objective. The results of the measurement are shown in Fig. 3(c,d) where it can be seen that minima that go to 0 in the tabletop case are greatly inflated when measured under the microscope due to the effect of now sampling the $d_{31}$, $d_{33}$ and $d_{15}$ coefficients as well. The magnitude of the component of the polarization that projects along the z-axis is polarization independent, and thus the contribution from these coefficients can be understood as a constant offset added on to the polar plots collected on the tabletop. A lack of uniformity in this offset as seen in the experimental data is attributed to subtle misalignment, since with $d_{33}$ as large as it is, any variation in the out-of-plane polarization component of the incident probe as the polarization is swept will be dramatically reflected in the generated SHG. In the next section, we provide numerical simulations for high NA polarimetry that provide better fits than the plane wave fits shown in Fig. 3.

### 4.2 Simulated Polarimetry Curves for High NA Objectives

***α-Quartz and LiNbO₃:*** The results of predicting the polarimetry curves discussed above using the methods described in section 3 are shown in Fig. S3. For both the *X* and *Y* polarized SHG curves of $\alpha$-quartz, very good agreement between theory and experiment is achieved. Theoretical curves are also able to capture the inflated minima of z-cut LiNbO₃ polarimetry curves due to the ability of a high NA lens to excite $P_3^{2\omega}$ through interaction with the $d_{33}$ coefficient and collect SHG emitted by $P_3^{2\omega}$ originating from interactions with $d_{31}$ and $d_{33}$. Additional high NA polarimetry measurements were performed on single crystal x-cut LiNbO₃, also purchased from MTI Corporation whose out-of-plane surface normal direction is the [11$\bar{2}$0]. In measuring x-cut LiNbO₃, the crystal physics axis 3 is in-plane along with projections of the 1 and 2 axes, providing greater access to the dominant $d_{33}$ coefficient and access to all tensor elements as a whole. In the *X*-polarized polarimetry curve, the non-zero minima along the 90° and 270° reflects interaction with $d_{31}$ and is non-zero even in the low NA case, while the maxima observed along 0° and 180° demonstrates the effect of having a dominant coefficient accessible with an in-plane polarization. In the *Y*-polarized polarimetry curve, the lobes near 135° and 315° reflect the magnitude of $d_{15}$, with divergence between simulated and experimental curves possibly suggesting a deviation from the assumed Kleinmann symmetry of the coefficients $d_{31} = d_{15}$, potentially due to the proximity of the measurement wavelength to the bandgap of the material.

The case study of z-cut LiNbO₃ best demonstrates the unique capability of performing polarimetry with high NA lenses, due to the detected signal being composed of nearly equal contributions from in-plane and out-of-plane nonlinear polarizations. A decomposition into these terms is highlighted in Fig. 4, with the in-plane contributions matching the plane-wave approximation solutions presented in Fig. 3 as can be expected. When the dimensions and orientation of the sample are constrained, as is commonly the case when investigating 2D or thin film materials, measurement with a high NA objective lens in this fashion may be used to access property tensor elements associated with an out-of-plane polarization developed in the material. By comparing different points in the polarimetry curve and interpreting them through these decompositions, measures identifying the ratios of SHG coefficients can be established with high accuracy.

***Al$_{0.94}$B$_{0.06}$N thin films***: A 20 nm Al$_{0.94}$B$_{0.06}$N (AlBN) thin film deposited on a 100 nm tungsten (W) on sapphire (Al$_2$O$_3$) substrate with the [0001] direction oriented out-of-plane was studied next. For details regarding the synthesis of the AlBN thin film, please see Ref. [34]. As a thin film on a metallic substrate, the AlBN sample is a departure from the previously measured single crystals but represents a case study where the out-of-plane polarization is the largest contributor to SHG signal. AlBN possesses the point group 6*mm*, with $d_{33}$ = 7.3 pm/V, $d_{31}$ = 0.35 pm/V, and $d_{15}$ = 0.7 pm/V for 6% B inclusion, resulting in the signal from $d_{33}$ being dominant if the incident probe polarization allows for such an interaction [34].

$$\text{Point group } 6mm: d_{ijk} = \begin{pmatrix} 0 & 0 & 0 & 0 & d_{15} & 0 \\ 0 & 0 & 0 & d_{15} & 0 & 0 \\ d_{31} & d_{31} & d_{33} & 0 & 0 & 0 \end{pmatrix} \quad (17)$$

As shown in Fig. S3(g,h), the polarimetry curves for AlBN show minimal dependence on incident polarization. This reflects again the fact that the probe component $E_3^\omega$ is independent of the incident polarization when the film is deposited with the [0001] direction out of the plane and confirms that the contribution from $P_3^{2\omega}$ to the overall signal dwarfs other contributions. The slight elliptical shape to the polarimetry curves for both polarized SHG components departing from the theoretical circular trace may indicate subtle changes in the reflection and transmission of p and s-polarized probe light through the experimental setup prior to sample interaction. Alternatively, they may reflect that the experimental probe profile is not truly Gaussian with cylindrical symmetry. Were the probe cross-section to be elongated along the lab Y direction, then a larger out-of-plane Z-component to the focused fields would be present when the polarization was aligned with the elongated axis. This would translate to a stronger SHG response for that input polarization, with the difference particularly apparent for a sample where the entire response relies upon the out-of-plane components, and with great sensitivity due to the quadratic dependence of the SHG response on the incident fields. While theoretical polarimetry curves were generated using an ideal Gaussian beam profile, experimental knife-edge measurement of the beam profile and corrected simulations yielding superior agreement are presented in Supplemental Fig. S4. The case study of [0001]-oriented AlBN also represents a material sample where symmetry would forbid the generation of any SHG signal when probed at normal incidence. In using a high NA objective however, sizeable SHG signal can be obtained and the polar nature of the material confirmed, arising predominantly from out-of-plane polarization.

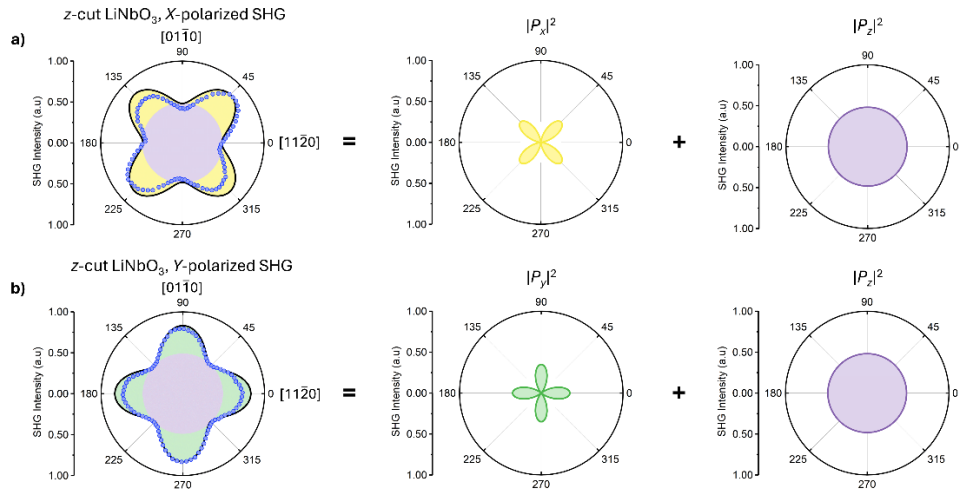

Fig. 4. Decomposition of simulated polarimetry curves for *z*-cut LiNbO$_3$ into individual contributions for in-plane (yellow or green) and out-of-plane (lavender) elements of the nonlinear polarization after transformation by the collecting lens.

## *4.3 Scanning SHG Microscopy*

With theoretical polarimetry curves yielding good agreement with experiment, it is possible to generate theoretical scanning SHG microscopy maps to predict the behavior of a sample. As an

SHG microscope generally performs signal detection by collecting all generated second harmonic light into a detector lacking spatial resolution, no traditional image of a sample is formed as is the case with an optical microscope. Instead, images are generated by raster-scanning the focused beam across a sample feature of interest, which can then be reconstructed as an area map. This can be handled using the method of section 2 by implementing a spatially varying form of the SHG tensor, provided one has an existing expectation for the phases and orientations of material present in a sample. As opposed to solving for a set of input polarizations, a fixed condition for the input polarization and analyzer orientation is chosen.

The fields described through $\vec{E}_{(ii)}(x, y, z)$ are discretely and numerically solved for a set of ($x,y,z$) points describing the focal plane or interaction volume and then used to evaluate the SHG equation, $P_i^{2\omega} = \varepsilon_o d_{ijk} E_j^\omega E_k^\omega$ to yield the emitted second harmonic fields $\vec{E}_{(iii)}$. In order to represent a multi-domain structure sampled by the probe beam, different combinations of $d_{ijk}$ are assigned to fields at different ($x,y,z$) coordinates and the final SHG signal fully evaluated. To simulate the effect of raster scanning the probe beam across such a feature, the condition used to describe the spatial variation of $d_{ijk}$ must be swept, for example by adding an offset in the $X$-direction at every step, and the final SHG signal evaluated at each step in the simulated raster scan. In using fully complex field quantities for evaluation, information regarding interference effects at domain boundaries will be naturally accounted for in the final SHG signal.

*Periodically poled LiNbO$_3$*: Using this method, a theoretical SHG map is generated for a standard example of periodically poled LiNbO$_3$ (PPLN) obtained from HC Photonics Corp. The bulk PPLN chip consists of inverted domains in an alternating, periodic fashion, such that the crystallographic [0001] is pointing out of the plane for one set of domains and the [000$\bar{1}$] direction for adjacent ones, referred to as up and down domains. The result is that from the frame of reference of the probe wave, the crystal physics 3 direction is inverted in one set of domains, effectively changing the sign of any $d_{ijk}$ SHG coefficient that has an odd combined number of subscripts 3 or 2. This change in sign of the coefficient results in a change in the sign of the amplitude of SHG generated from interaction with that coefficient, for example changing the sign of $P_i^{2\omega}$ across up and down domains when exciting through $d_{311}\equiv d_{31}$, $d_{333}\equiv d_{33}$, $d_{222}\equiv d_{22}$, $d_{113}\equiv d_{15}$, and $d_{223}\equiv d_{24}$. If the collimating lens collects SHG generated from adjacent domains in such a case, namely by being positioned above a domain wall, the SHG signal from the two domains will destructively interfere and a reduction in SHG signal will be observed at the domain wall. This phenomenon is captured in the experimental data shown in Fig. 5, and accurately simulated by implementing a spatial distribution of the two different types of SHG coefficient tensors as described above. The observed difference in signal between one set of domains and the other is attributed to a step-like height differential of roughly 340 nm observed at the PPLN domain boundaries using atomic force microscopy (AFM), a discussion on which is provided in supplemental Fig. S5 and the accompanying text. This step-like surface feature is also expected to introduce scattering and depolarization of second harmonic light emitted by the sample, reducing the ability of light generated from adjacent domains to completely interfere destructively. Exotic domain wall structures possessing SHG active local symmetry may emerge in certain materials and can be uniquely detected using SHG microscopy. For a focused discussion on such complex structures, readers are directed to [19].

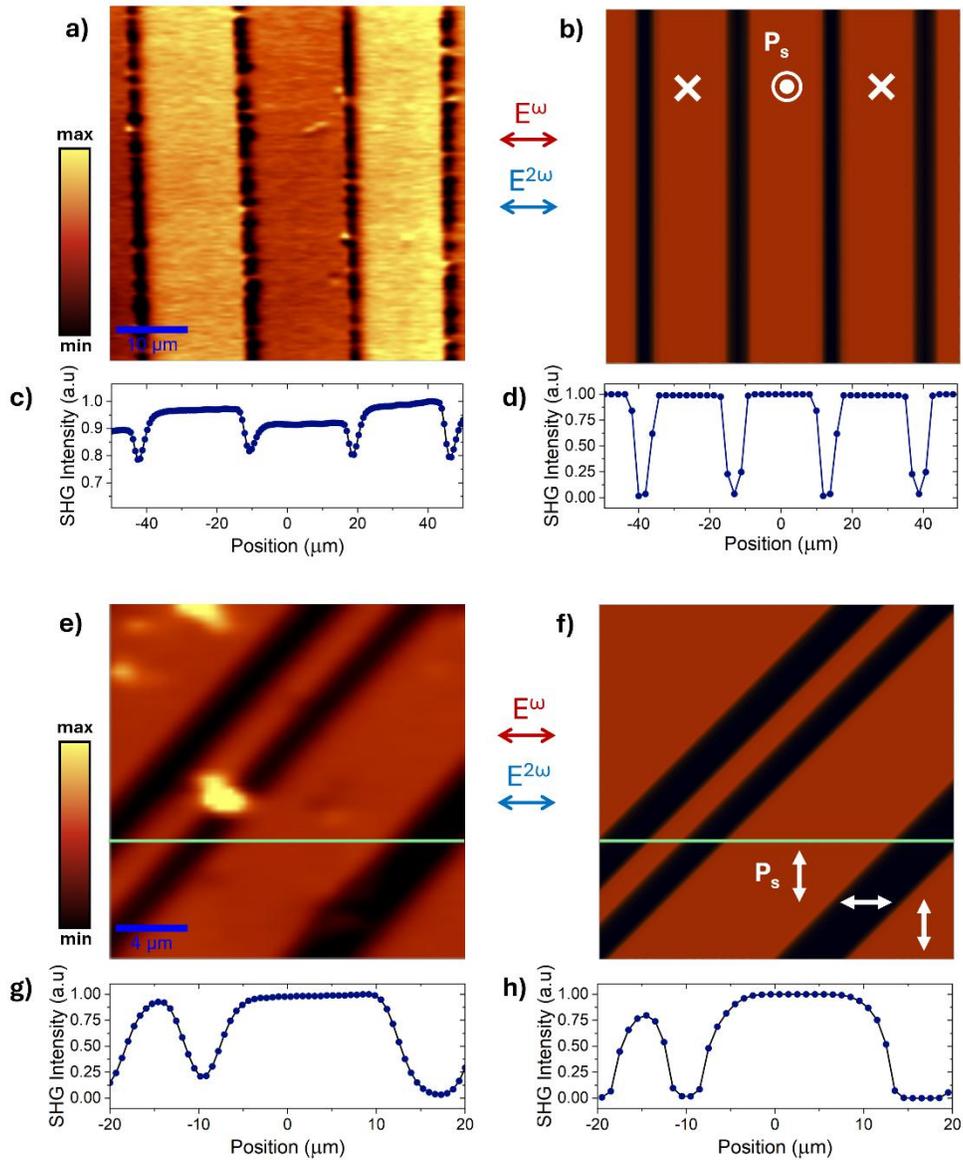

Fig. 5. a) Experimentally obtained SHG scanning map on periodically poled LiNbO$_3$, with line cut of average signal shown in (c). b) Theoretically predicted area map of similar structure with representative line scan in (d). e) Experimentally obtained SHG scanning map on natural *a*1 and *a*2 domains in a BaTiO$_3$ single crystal, with a line cut taken from the line indicated in green shown in (g). Theoretically predicted area map of similar structure with representative line scan in (h). Arrows alongside P$_s$ indicate the direction of the polar axis in the different domains of both structures.

***BaTiO$_3$ single crystal***: A similar SHG microscopy map is taken from a BaTiO$_3$ single crystal sample with point group 4*mm*, which possesses natural domains for two different orientations of the in-plane polarization, termed *a*1 and *a*2 domains. In one set of domains, the polar axis is aligned along the lab *X* axis, while in the other it is aligned along the lab *Y*. This results in a simple case where a polarization aligned along *X* projects along the crystal physics 3 axis in one domain, and the crystal physics 1 axis in the other. In performing the appropriate coordinate transforms for each case on the property tensor shown below, a simulated map is generated to mimic the experimentally observed structure.

$$\text{Point group } 4mm: d_{ijk} = \begin{pmatrix} 0 & 0 & 0 & 0 & d_{15} & 0 \\ 0 & 0 & 0 & d_{15} & 0 & 0 \\ d_{31} & d_{31} & d_{33} & 0 & 0 & 0 \end{pmatrix} \quad (18)$$

As opposed to the case of periodically poled $LiNbO_3$, dark regions are no longer due to interference at a domain wall, but a reflection of the fact that the selected combination of incoming polarization and analyzer orientation results in no measured SHG signal from that set of domains. Polarimetry curves and simulated fits collected for each type of domain are provided in supplementary Fig. S6.

*4.4 The Inverse Problem: Extracting SHG Coefficients from Polarimetry Maps*
Thus far, we have demonstrated that for well-known crystals with known SHG tensor coefficients, the predicted polarimetry from the above theoretical approach is in good agreement with experiments. The "inverse problem" here would be to find the SHG tensor for a new material starting from the experimental polarimetry data. There are two aspects to extracting the SHG tensor: (1) finding the ratios of various SHG coefficients, which can be obtained from the obtained shapes of the polar plots, and (2) finding the absolute values of the SHG coefficients, which requires a standard measurement to compare against. We discuss each of these aspects next.

The modeling framework presented in this work relies upon numerical solutions for final SHG intensities for speed and ease of computation. As such, retrieval of analytical solutions for fitting experimental data must be approached deliberately. One can focus on critical points within polarimetry curves – for example maxima and minima – and vary the magnitude of one tensor coefficient while setting all others to 0. In this way, a quadratic dependence of the final integrated intensity on that single element can be obtained, with a leading scalar reflecting other modelling parameters such as the chosen input beam profile, the NA of the lens, and the refractive index of the material. The process can then be repeated for all elements in the property tensor, with the final intensity being a sum of all such terms. Example fitting equations for $z$-cut $LiNbO_3$ and a more detailed discussion of the fitting procedure is presented in Supplemental 1. The results of retrieving ratios of SHG coefficients at a fundamental wavelength of 800 nm for several of the experimental measurements collected within this work are presented below in Table 1.

**$LiNbO_3$**: The ratios of SHG coefficients, $d_{33}/d_{31}$, $d_{33}/d_{15}$, and $d_{33}/d_{22}$ were determined using both $x$-cut and $z$-cut $LiNbO_3$ where $x$ and $z$ refer to the crystallographic $[11\bar{2}0]$ and $[0001]$ directions respectively as in the previous sections. Kleinmann symmetry was assumed to be true for the property tensor of $LiNbO_3$: $d_{31} = d_{15}$. The experimentally determined ratios and the values from the literature are in excellent agreement, thus providing confidence in our approach.

Quantitative extraction of SHG coefficients requires a reliable standard with which to analyze the high NA polarimetry data. In our case, the standard was established using a wedged crystal of $z$-cut quartz or an $x$-cut $LiNbO_3$ (identical to those described in previous sections) whose surfaces were probed using a 0.12 NA lens in a normal reflection geometry. Polishing the crystal to a wedge eliminates reflections from the back surface thus greatly reducing the complications arising from multiple reflections from the two interfaces. The lower NA used for the standard measurement ensures that the absolute value determined for the standard can use the plane-wave approximation. Once a single coefficient is determined as a standard in this manner, the ratios of the SHG coefficients already determined from the high NA polarimetry data are utilized to determine the rest of the coefficients.

Alignment of the incoming polarization and analyzer along the [0001] direction of LiNbO$_3$ allows for pure sampling of the $d_{33}$ coefficient as a standard measurement. Similarly, aligning the optics with the [11$\bar{2}$0] direction of quartz isolates and determines $d_{11}$ as a standard. By modeling the SHG intensities collected at these critical conditions, their respective tensor elements can be directly compared. Using quartz as the reference material with a value of $d_{11}$ = 0.3 pm/V provided by literature, the value of $d_{33}$ in LiNbO$_3$ was quantified at 27.5 ± 0.1 pm/V, in close agreement with the literature values [9,33]. By combining a low NA measurement designed to quantity a single coefficient with the ability to sample multiple coefficients and retrieve their ratios with a single high NA measurement, one can then proceed with quantifying the entire material property tensor.

**BaTiO$_3$**: The ratios $d_{33}/d_{31}$ and $d_{33}/d_{15}$ for BaTiO$_3$ were also determined from the SHG polarimetry, where $d_{31}$ and $d_{15}$ were allowed to be unequal due to non-zero extinction coefficients at the wavelengths of measurement. Here, the agreement with literature ratios is less satisfactory, with a disagreement of ~25% for $d_{33}/d_{15}$ and up to ~57% for $d_{33}/d_{31}$. The reason for these disagreements arises from the fact that SHG ratios in polarimetry are often related to ratios of various intensity *minima* in the polar plots at different fundamental and SHG polarization orientations. These minima can be easily influenced by slight crystal misalignments with respect to the optics, or the scattering of light from internal interfaces and rough surfaces resulting in some depolarization. In this work, such effects were observed in measuring polarimetry curves from domains in a BaTiO$_3$ single crystal shown in supplementary Fig. S5, resulting in the observed deviations in ratios given in Table 1.

**Table 1. Extracted SHG Coefficient Ratios from Benchmark Materials at 800 nm Fundamental Wavelength**

|  | This Work | | | Literature [a] | | |
|---|---|---|---|---|---|---|
| Sample | $d_{33}/d_{31}$ | $d_{33}/d_{15}$ | $d_{33}/d_{22}$ | $d_{33}/d_{31}$ | $d_{33}/d_{15}$ | $d_{33}/d_{22}$ |
| z-cut LiNbO$_3$[b] | 6.13±0.06 | 1 | 13.21±0.21 | 6.25 | 1 | 12.95 |
| x-cut LiNbO$_3$[b] | 3.82±0.27 | 1 | 11.77±0.12 | 6.25 | 1 | 12.95 |
| BaTiO$_3$[c] | 0.60±0.01 | 0.50±0.01 | | 0.38 | 0.40 | |

| Sample | $d_{11}/d_{22}$ | $d_{11}/d_{31}$ | $d_{11}/d_{33}$ |
|---|---|---|---|
| SnP$_2$Se$_6$[d] | 1.88 ± 0.50 | 2.74 ± 0.79 | 0.04 ± 0.01 |

[a]From [33]; [b]$d_{33}$ = 27.5 ± 0.1 pm/V from 0.12 NA measurement; [c]$d_{31}$ = 14.4 ± 1.2 pm/V from [33]; [d]$d_{11}$ = 15.1 ± 0.2 pm/V from 0.12 NA measurement

**SnP$_2$Se$_6$**: As a demonstration of the ability to apply this to a less well-studied material, full polarimetry measurements were performed with low and high NA lenses on crystals of SnP$_2$Se$_6$ at a fundamental wavelength of 800 nm. SnP$_2$Se$_6$ is van der Waals semiconductor that has been noted to have strong optical nonlinearities at 1550 nm in the 2D limit [35–37]. An initial guess for the crystal symmetry is recommended to be known beforehand from X-ray or neutron diffraction, which can provide both the highest possible symmetry group as well as the sample orientation. SnP$_2$Se$_6$ has a point group symmetry of 3, for which the SHG tensor is given as follows:

$$\text{Point group 3: } d_{ijk} = \begin{pmatrix} d_{11} & -d_{11} & 0 & d_{14} & d_{15} & -d_{22} \\ 0 & 0 & 0 & d_{15} & -d_{14} & -d_{11} \\ d_{31} & d_{31} & d_{33} & 0 & 0 & 0 \end{pmatrix} \quad (19)$$

The results of the polarimetry experiments are shown in Fig. 6. The crystal physics 1 axis is parallel to the crystallographic [01$\bar{1}$0], the crystal physics 3 axis parallel with the crystallographic [0001], and the crystal physics 2 along 3 × 1. The response is dominated by the $d_{11}$ and $d_{22}$ coefficients that are directly probed by the in-plane polarized component of the beam. Under high NA, an inflation of the minima is observed associated with probing the $d_{31}$ and $d_{33}$ coefficients with the non-zero out-of-plane projection of the polarization. Owing to a bulk bandgap of 1.52 eV, both the fundamental and second harmonic wavelengths are expected to be strongly absorbed, fulfilling the single interface SHG condition that enables referencing of the low NA measurement against a wedged quartz sample. In doing so, $d_{11}$ is quantified as 15.1 ± 0.2 pm/V, while fitting of the high NA data yields coefficient ratios of $d_{11}/d_{22}$ = 3.53 ± 0.76, $d_{11}/d_{31}$ = 2.27 ± 0.85, and $d_{11}/d_{33}$ = 0.07 ± 0.01. The $d_{14}$ and $d_{15}$ coefficients are unable to be determined in this geometry due to cylindrical symmetry of the focused Gaussian probe.

This phenomenon can be understood by considering the relative sign of the fields within the focal region. For measurement of a material of point group 3 with a probe propagating along the 3 axis, $d_{14}$ and $d_{15}$ can be seen to manifest in $P_1$ as terms that will go $2d_{14}E_2E_3$ and $2d_{15}E_1E_3$ respectively with similar terms existing in $P_2$. Within the focused beam profile, the sign of $E_2$ and $E_3$ components will depend on the position of a given ray with respect to the plane containing the optical axis of the focusing lens and perpendicular to the incident polarization (assuming the 2 and 3 directions are aligned with the Lab Y and Lab Z axes respectively). This aforementioned plane can be seen as a mirror plane in the focused intensities of Fig. 1(c-e) but will not serve as a mirror plane for the focused $E_2$ and $E_3$ electric field quantities. As such, for any given ray within the beam profile, there will exist a mirrored counterpart where $E_1$ is of the same sign, but $E_2$ and $E_3$ are opposite. Therefore, the terms dependent on $d_{14}$ and $d_{15}$ will have counterparts equal in intensity but opposite in sign once the entire beam profile is integrated over, resulting in no dependence on $d_{14}$ and $d_{15}$ in the final SHG intensity assuming the center of the Gaussian probe is aligned with the optical axis of the focusing lens.

This aspect reveals a limitation of this approach. More generally, the spatial profiles of the focused $E_Y$ and $E_Z$ fields do not possess mirror symmetry. As a result, contributions from $d_{ijk}$ tensor elements where $j, k$ are odd in Y or Z will not provide contributions to measured SHG intensities due to the aforementioned cancellation when the final integration over the beam profile is performed. This assumes that the crystal physics axes are aligned with the lab axes as is generally recommended. Intentional misalignment of the probe beam, namely offsetting the center of a Gaussian profile from the center of the lens, could serve to break this symmetry. However, such a measure will also render the aplanatic condition of the system invalid and potentially introduce aberration. For small deviations from the aplanatic condition, the current approach appears to be in reasonable agreement with finite element method COMSOL numerical solutions included in Fig. S1.

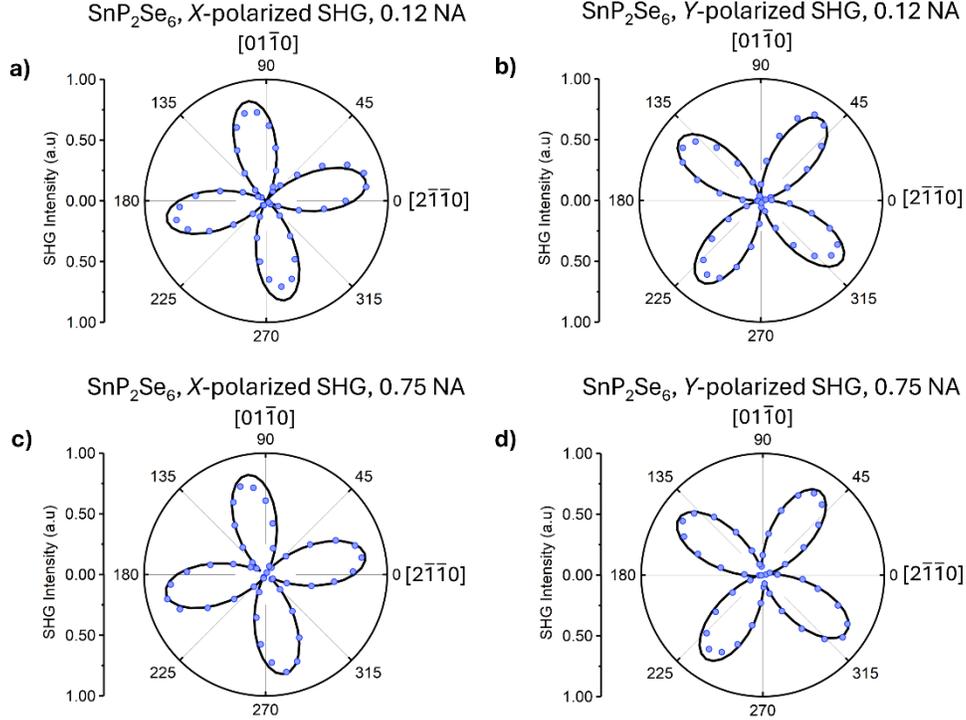

Fig. 6. Experimentally obtained (dotted, blue) and simulated (solid, black) SHG polarimetry curves obtained from single crystal SnP$_2$Se$_6$ with a 0.12 NA (a,b) and 0.75 NA (c,d) focusing lens. The *X*-direction of the setup is aligned with the [2$\bar{1}\bar{1}$0] crystal direction while the [0001] is aligned with the out-of-plane *Z*-direction of the setup.

Using the above analysis, we can now extract the spatial map of the SHG coefficients for SnP$_2$Se$_6$. This is shown in a series of images in Fig. 7. The origin of these significant spatial variations in the SHG tensor can be understood from analyzing the complete polarimetry data in each distinct region. Polarimetry from four distinct regions labeled I-IV in Fig. 8(a) are shown in Fig. 8(b). From the analysis, we conclude that regions 1-3 all have the same orientation of their crystal physics axes though their SHG intensities vary considerably. This indicates that intensity variations in the maps in Fig. 7 arise from inversion domains as schematically shown in Fig. 8(c), where different layers of this material in the thickness direction have a reversal of all three crystal physics axes, namely, $1 \rightarrow -1, 2 \rightarrow -2, 3 \rightarrow -3$, and hence $d_{ijk} \rightarrow -d_{ijk}$. Depending upon the volume fraction difference between the inverted domains within the probe depth (11.7 μm for a probe at 800 nm), one can get different intensities in regions 1-3, though their polarimetry remains the same. In this sense, the $d_{ijk}$ maps shown in Fig. 7 should be considered "effective" coefficients, where local domain structure influences the absolute value of these coefficients. Region 4 appears distinct from regions 1-3 as shown in Fig. 8(b) where its polarimetry data is rotated by ~4°. These appear to be in-plane rotated domains with the tensor relationship between regions labeled 1 and 4 being: $1 \rightarrow 1, 2 \rightarrow -2, 3 \rightarrow -3$ as illustrated in Fig. 8(d). Our work thus indicates that these types of subtle domain orientation distinctions as well as the effective nonlinear coefficients within the probe depth can be qualitatively and quantitatively extracted from high NA SHG microscopy of a previously unknown material, thus illustrating our ability to address the stated inverse problem.

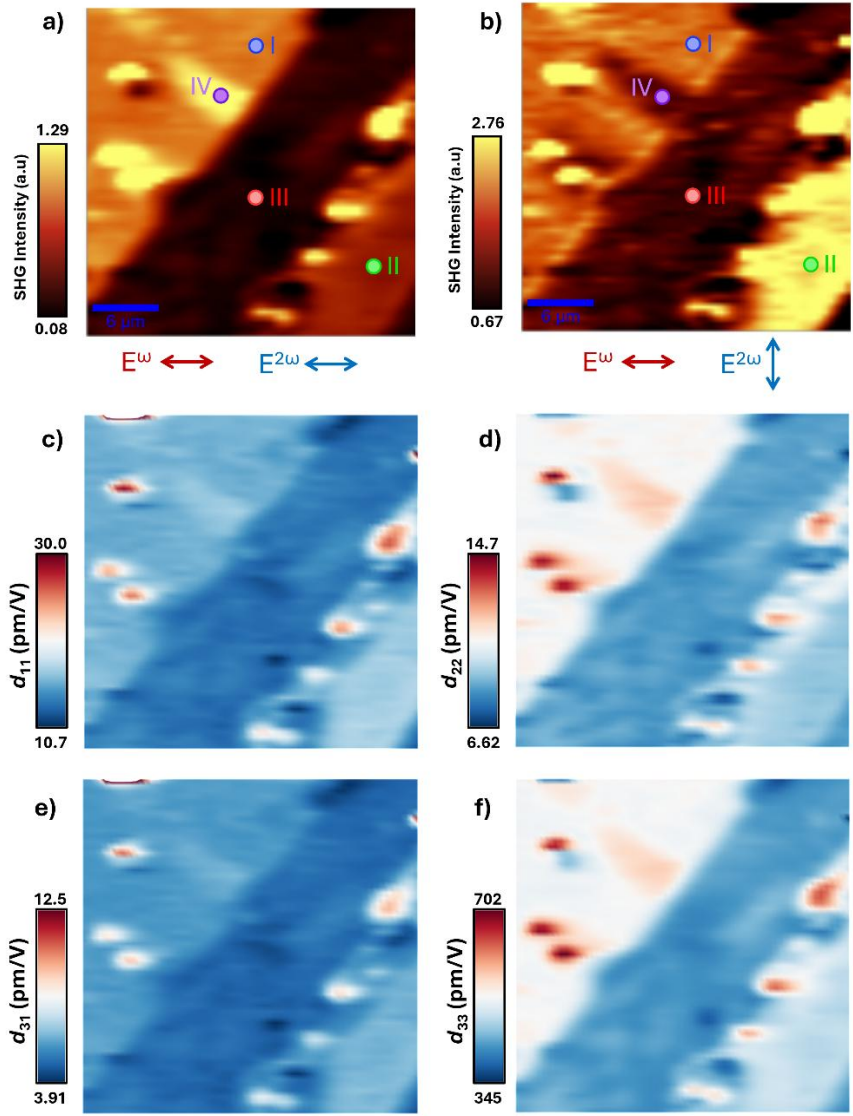

Fig. 7. Experimentally obtained SHG intensity maps of SnP$_2$Se$_6$ obtained at two critical configurations for coefficient ratio extraction: lab $X$ fundamental polarization in with a) lab $X$ or b) lab $Y$ SHG polarization detected. The $X$-direction of the setup is aligned with the $[2\bar{1}\bar{1}0]$ crystal direction and the $Y$-direction aligned with $[01\bar{1}0]$. c-f) SHG coefficient maps obtained by collecting maps at additional critical points and performing simultaneous point-by-point fits to obtain the location dependence of property tensor coefficient ratios. The average value of the $d_{11}$ coefficient map was set to the value obtained via 0.12 NA measurements and used to calibrate the values of other coefficients.

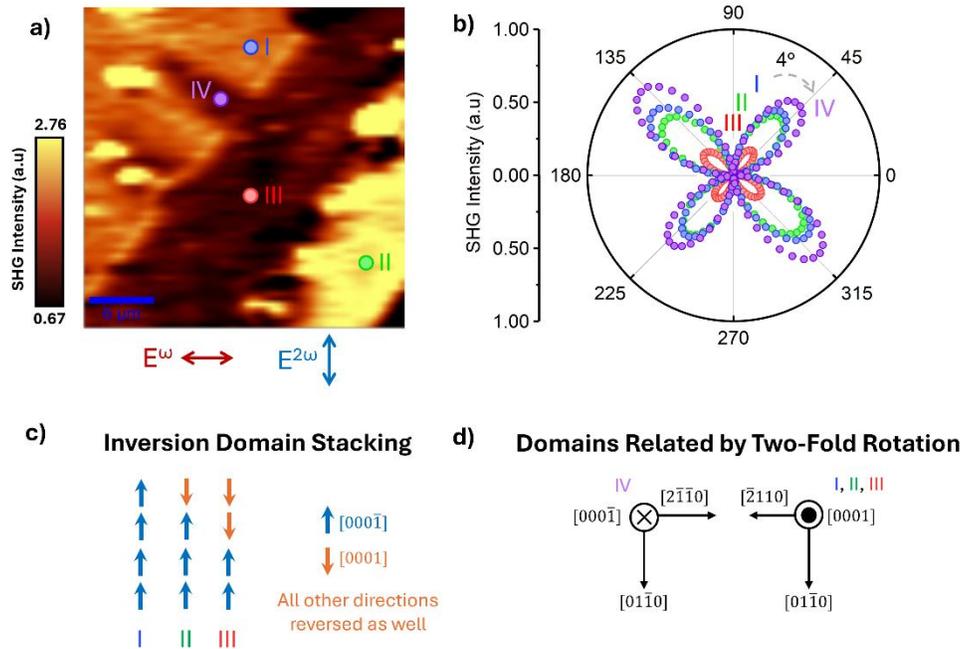

Fig. 8. a) Experimentally obtained SHG intensity map of $SnP_2Se_6$ obtained with lab $X$ fundamental polarization in and lab $Y$ SHG polarization detected. The $X$-direction of the setup is aligned with the $[2\bar{1}\bar{1}0]$ crystal direction and the $Y$-direction aligned with $[01\bar{1}0]$. b) Local SHG polarimetry performed at regions of interest indicated with colored circles in (a). c) Schematic of an example stacking arrangement of inversion domain variants potentially responsible for reduction of SHG signal while preserving shape of polarimetry. d) Schematic of domain variants related by two-fold rotation about $[01\bar{1}0]$. The crystal physics 1 direction is preserved, preventing full signal cancellation and instead rotating the polarimetry.

## 5. Conclusions

A method for quantitatively predicting second harmonic generation polarimetry data collected when using high NA lenses is developed. Such an approach is necessary due to the failure of the paraxial approximation when the lenses used focus the probe beam very tightly, resulting in a complex polarization scheme within the focal region. Since the SHG process is highly polarization sensitive, attention must be given to solving the fields within the interaction volume, and then using the focused fields to generate nonlinear dipole sources of second harmonic radiation. High NA lenses are capable not only of projecting the polarization of the probe along the out-of-plane direction, but also collecting second harmonic signal emitted from nonlinear dipole sources aligned along the out-of-plane direction too. By performing an inverse transformation of the polarization of the second harmonic fields analogous to that used to represent the focusing of the probe fields, and then integrating over the collection angle of the lens, the final second harmonic signal is captured including the contribution from the out-of-plane component. The result is the ability to sample SHG coefficients that are unable to be accessed in traditional geometries under the paraxial approximation.

Theoretical curves generated using this method provide good agreement with experiment for a variety of test cases that demonstrate the range of features that can manifest when having no out-of-plane polarization component, to a dominant out-of-plane polarization. By predicting the SHG signal generated at a fixed polarization for a spatially varying material property tensor, it is also possible to simulate a scanning SHG microscope experiment where complex domain structures are imaged. Interaction between adjacent domains is accurately captured, allowing for modeling of the behavior of domain wall features. A solution to the inverse problem is demonstrated, where the unknown SHG tensor of a crystal is determined by experimentally measured polarimetry. This work focuses on a single sample interface; future extensions of the work could involve modeling multiple reflections of both the fundamental and the SHG waves within multilayer samples. With significant interest in the study of 2D materials using SHG microscopy, picking out out-of-plane components with SHG microscopy becomes of significant value [38]. This work represents a step forward towards quantitative SHG microscopy and polarimetry performed using Gaussian beams focused through high NA objectives.

**Funding.** Basic Energy Sciences (DE-SC0021118); Air Force Office of Scientific Research (FA 9550-23-1-0658); Basic Energy Sciences (DE-SC0012375)

**Acknowledgments.** The optical characterization and experimental modelling performed by A.S. and AlBN thin film growth performed by J.H. are based upon work supported by the center for 3D Ferroelectric Microelectronics (3DFeM2), an Energy Frontier Research Center funded by the U.S. Department of Energy, Office of Science, Office of Basic Energy Sciences Energy Frontier Research Centers program under Award Number DE-SC0021118. $SnP_2Se_6$ crystal growth performed by S.I. and A.I. and structural characterization of these crystals performed by J.N. were supported by the Air Force Office of Scientific Research through the Grant award number FA 9550-23-1-0658. Atomic force microscopy data collected by S.H. was supported by the U.S. Department of Energy, Office of Science, Office of Basic Energy Sciences Award Number DE-SC0012375. The authors would also like to acknowledge Rui Zu, Akash Saha, and Saugata Sarker for their discussions on this work.

**Disclosures.** All authors declare no conflict of interest.

**Data availability.** Data underlying the results presented in this paper are not publicly available at this time but may be obtained from the authors upon reasonable request.

**Supplemental document.** See Supplement 1 for supporting content.

**References**

1. S. A. Denev, T. T. A. Lummen, E. Barnes, A. Kumar, and V. Gopalan, "Probing ferroelectrics using optical second harmonic generation," Journal of the American Ceramic Society **94**(9), 2699–2727 (2011).
2. F. Simon, S. Clevers, V. Dupray, and G. Coquerel, "Relevance of the Second Harmonic Generation to Characterize Crystalline Samples," Chem Eng Technol **38**(6), 971–983 (2015).
3. Y. Wang, J. Xiao, S. Yang, Y. Wang, and X. Zhang, "Second harmonic generation spectroscopy on two-dimensional materials," Opt. Mater. Express **9**(3), 1136–1149 (2019).
4. K. H. Matlack, J.-Y. Kim, L. J. Jacobs, and J. Qu, "Review of Second Harmonic Generation Measurement Techniques for Material State Determination in Metals," J Nondestr Eval **34**(1), 273 (2014).
5. Y.-R. Shen, "Optical second harmonic generation at interfaces," Annu Rev Phys Chem **40**(1), 327–350 (1989).
6. P. Campagnola, "Second Harmonic Generation Imaging Microscopy: Applications to Diseases Diagnostics," Anal Chem **83**(9), 3224–3231 (2011).
7. R. J. Tran, K. L. Sly, and J. C. Conboy, "Applications of Surface Second Harmonic Generation in Biological Sensing," Annual Review of Analytical Chemistry **10**(Volume 10, 2017), 387–414 (2017).
8. P. J. Campagnola, A. C. Millard, M. Terasaki, P. E. Hoppe, C. J. Malone, and W. A. Mohler, "Three-Dimensional High-Resolution Second-Harmonic Generation Imaging of Endogenous Structural Proteins in Biological Tissues," Biophys J **82**(1), 493–508 (2002).
9. Robert W. Boyd, *Nonlinear Optics* (Elsevier, 2003).
10. R. Newnham, *Properties of Materials: Anisotropy, Symmetry, Structure* (Oxford University Press, 2005).
11. W. N. Herman and L. M. Hayden, "Maker fringes revisited: second-harmonic generation from birefringent or absorbing materials," J. Opt. Soc. Am. B **12**(3), 416–427 (1995).
12. R. Zu, B. Wang, J. He, J. J. Wang, L. Weber, L. Q. Chen, and V. Gopalan, "Analytical and numerical modeling of optical second harmonic generation in anisotropic crystals using ♯SHAARP package," NPJ Comput Mater **8**(1), (2022).
13. N. Bloembergen and P. S. Pershan, "Light Waves at the Boundary of Nonlinear Media," Physical Review **128**(2), 606–622 (1962).
14. I. Shoji, T. Kondo, A. Kitamoto, M. Shirane, and R. Ito, "Absolute scale of second-order nonlinear-optical coefficients," J. Opt. Soc. Am. B **14**(9), 2268–2294 (1997).

15. R. Zu, B. Wang, J. He, L. Weber, A. Saha, L.-Q. Chen, and V. Gopalan, "Optical second harmonic generation in anisotropic multilayers with complete multireflection of linear and nonlinear waves using ♯SHAARP.ml package," NPJ Comput Mater **10**(1), 64 (2024).
16. S. Trolier-McKinstry and R. E. Newnham, *Materials Engineering: Bonding, Structure, and Structure-Property Relationships* (Cambridge University Press, 2017).
17. X. H. Wang, S. J. Chang, L. Lin, L. R. Wang, B. Z. Huo, and S. J. Hao, "Vector model for polarized second-harmonic generation microscopy under high numerical aperture," Journal of Optics A: Pure and Applied Optics **12**(4), (2010).
18. B. Richards, E. Wolf, and D. Gabor, "Electromagnetic diffraction in optical systems, II. Structure of the image field in an aplanatic system," Proc R Soc Lond A Math Phys Sci **253**(1274), 358–379 (1959).
19. Y. Zhang and S. Cherifi-Hertel, "Focusing characteristics of polarized second-harmonic emission at non-Ising polar domain walls," Opt Mater Express **11**(11), 3736–3754 (2021).
20. M. Leutenegger, R. Rao, R. A. Leitgeb, and T. Lasser, "Fast focus field calculations," Opt. Express **14**(23), 11277–11291 (2006).
21. M. Strupler, A.-M. Pena, M. Hernest, P.-L. Tharaux, J.-L. Martin, E. Beaurepaire, and M.-C. Schanne-Klein, "Second harmonic imaging and scoring of collagen in fibrotic tissues," Opt Express **15**(7), 4054–4065 (2007).
22. S.-J. Lin, C.-Y. Hsiao, Y. Sun, W. Lo, W.-C. Lin, G.-J. Jan, S.-H. Jee, and C.-Y. Dong, "Monitoring the thermally induced structural transitions of collagen by use of second-harmonic generation microscopy," Opt Lett **30**(6), 622–624 (2005).
23. C.-L. Hsieh, Y. Pu, R. Grange, and D. Psaltis, "Second harmonic generation from nanocrystals under linearly and circularly polarized excitations," Opt Express **18**(11), 11917–11932 (2010).
24. H. Yokota, T. Hayashida, D. Kitahara, and T. Kimura, "Three-dimensional imaging of ferroaxial domains using circularly polarized second harmonic generation microscopy," NPJ Quantum Mater **7**(1), 106 (2022).
25. P. Török, P. D. Higdon, and T. Wilson, "Theory for confocal and conventional microscopes imaging small dielectric scatterers," J Mod Opt **45**(8), 1681–1698 (1998).
26. V. Kopský and D. B. Litvin, eds., *International Tables for Crystallography* (International Union of Crystallography, 2006), **E**.
27. M. O. Ramirez, T. T. A. Lummen, I. Carrasco, E. Barnes, U. Aschauer, D. Stefanska, A. Sen Gupta, C. de las Heras, H. Akamatsu, M. Holt, P. Molina, A. Barnes, R. C. Haislmaier, P. J. Deren, C. Prieto, L. E. Bausá, N. A. Spaldin, and V. Gopalan, "Emergent room temperature polar phase in CaTiO3 nanoparticles and single crystals," APL Mater **7**(1), 011103 (2019).
28. E. N. Economou, *Green's Functions in Quantum Physics* (Springer Berlin Heidelberg, 1983), **7**.
29. A. V. Lavrinenko, J. Lægsgaard, N. Gregersen, F. Schmidt, and T. Søndergaard, *Numerical Methods in Photonics* (CRC Press, 2018).
30. T. Søndergaard, "Modeling of plasmonic nanostructures: Green's function integral equation methods," physica status solidi (b) **244**(10), 3448–3462 (2007).
31. O. J. F. Martin, A. Dereux, and C. Girard, "Iterative scheme for computing exactly the total field propagating in dielectric structures of arbitrary shape," Journal of the Optical Society of America A **11**(3), 1073–1080 (1994).
32. A. Teulle, R. Marty, S. Viarbitskaya, A. Arbouet, E. Dujardin, C. Girard, and G. Colas des Francs, "Scanning optical microscopy modeling in nanoplasmonics," Journal of the Optical Society of America B **29**(9), 2431–2437 (2012).
33. D. N. Nikogosyan, *Nonlinear Optical Crystals: A Complete Survey* (Springer-Verlag, 2005).
34. A. Suceava, J. Hayden, K. P. Kelley, Y. Xiong, B. Fazlioglu-Yalcin, I. Dabo, S. Trolier-McKinstry, J.-P. Maria, and V. Gopalan, "Enhancement of second-order optical nonlinearities and nanoscale periodic domain patterning in ferroelectric boron-substituted aluminum nitride thin films," Opt. Mater. Express **13**(6), 1522–1534 (2023).
35. G. M. Maragkakis, S. Psilodimitrakopoulos, L. Mouchliadis, A. S. Sarkar, A. Lemonis, G. Kioseoglou, and E. Stratakis, "Nonlinear Optical Imaging of In-Plane Anisotropy in Two-Dimensional SnS," Adv Opt Mater **10**(10), 2102776 (2022).
36. V. K. Sangwan, D. G. Chica, T.-C. Chu, M. Cheng, M. A. Quintero, S. Hao, C. E. Mead, H. Choi, R. Zu, J. Sheoran, J. He, Y. Liu, E. Qian, C. C. Laing, M.-A. Kang, V. Gopalan, C. Wolverton, V. P. Dravid, L. J. Lauhon, M. C. Hersam, and M. G. Kanatzidis, "Bulk photovoltaic effect and high mobility in the polar 2D semiconductor SnP2Se6," Sci Adv **10**(31), eado8272 (2024).
37. J. Nag, S. Sarker, S. Imam, A. Iyer, M. J. Waters, A. Suceava, J. M. Rondinelli, M. G. Kanatzidis, and V. Gopalan, "Large Non-Resonant Infrared Optical Second Harmonic Generation in Bulk Crystals of Van der Waals Semiconductor, SnP2Se6," Adv Opt Mater **13**(9), 2402649 (2025).
38. H. Ma, J. Liang, H. Hong, K. Liu, D. Zou, M. Wu, and K. Liu, "Rich information on 2D materials revealed by optical second harmonic generation," Nanoscale **12**(45), 22891–22903 (2020).

# Quantitative Nonlinear Optical Polarimetry with High Spatial Resolution

Albert Suceava[1], Sankalpa Hazra[1], Jadupati Nag[1], John Hayden[1], Safdar Imam[2], Zhiwen Liu[3], Abishek Iyer[2], Mercouri Kanatzidis[2], Susan Trolier-McKinstry[1], Jon-Paul Maria[1], Venkatraman Gopalan[1, 4, 5]*

[1] Department of Materials Science and Engineering and the Materials Research Institute, The Pennsylvania State University, University Park, Pennsylvania 16802, USA
[2] Department of Chemistry, Northwestern University, Evanston, Illinois 60208, USA
[3] Department of Electrical Engineering, Pennsylvania State University, University Park, Pennsylvania, 16802, USA
[4] Department of Physics, Pennsylvania State University, University Park, Pennsylvania, 16802, USA
[5] Department of Engineering Science and Mechanics, Pennsylvania State University, University Park, Pennsylvania, 16802, USA
*vxg8@psu.edu

## Comparisons of Near Field Profile to Beam Envelopes Method

The plane-wave spectrum approach adopted from Ref. [1] assumes Abbe's sine condition, which precludes modelling the effects of common experimental nonidealities, like decentering of the probe with respect to the objective. Nonetheless, the flexibility in defining the input beam profile invites exploration of the accuracy of this approach against other computational tools. We show below that underfilling and non-centering of the beam on the objective lens results in fields that are in reasonable agreement between this work and COMSOL simulations.

The methodology of this work does not directly construct the spatial profile of a practical beam waist. Rather, the deflection of probe rays at the near field towards a focal point is calculated, and then propagation evaluated using the Debye diffraction integral. To serve as a point of comparison then, the near field spatial field profiles obtained in this framework are compared to analogous spatial profiles obtained using the beam envelopes method of COMSOL Multiphysics.

Specifically, a single spherical lens with index $n = 1.5$, diameter $D = 200$ μm, and radius of curvature $R = 100$ μm is constructed in a two-dimensional environment in COMSOL, as shown in Fig. S1(a). An input field profile of the form $\vec{E} = \exp[-r^2/2\sigma^2]\,\hat{x}$ is made incident on the lens using a Matched Boundary Condition, where the $\hat{x}$ direction is out-of-plane in the two-dimensional environment (analogous to the lab X direction of Fig. 1(a)). Perfectly matched layers are constructed on all other external boundaries to simulate open boundaries. An analogous lens dimension and input field profile is utilized with the methodology of this work and used to generate the near field spatial profile of the probe. Comparison between the COMSOL beam envelopes method and the method of this work is shown in Fig. S1, demonstrating good qualitative agreement in the near field for the case of decentered Gaussian profiles or underfilled apertures.

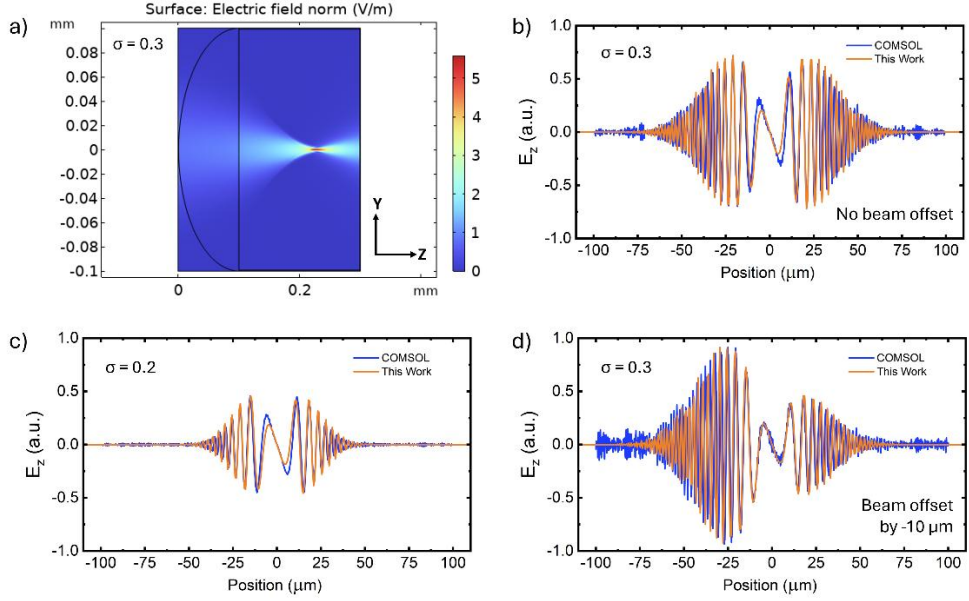

Fig. S1. a) Two-dimensional COMSOL Beam Envelopes Method simulation of a focused Gaussian probe of cross-sectional form $\vec{E} = \exp[-r^2/2\sigma^2]$, polarized out-of-plane, along lab X. Comparison of the electric field components in the near field along the direction of propagation, Z, for (b) $\sigma = 0.3$ and (c) $\sigma = 0.2$ between COMSOL and the methodology of this work. d) Comparison between COMSOL and this work for a Gaussian beam with $\sigma = 0.3$ but with origin offset by 10 μm in the -Y direction.

## Generalization of the Approach to Higher Order Nonlinear Processes

The approach established in this work can be thought of as having two aspects: firstly, the generation of focal fields under tight focusing conditions and secondly, the use of those fields to model the nonlinear optical response. For other nonlinear optical processes aside from second harmonic generation (SHG), the framework can be extended by adapting the second half (from $\vec{E}_{(iv)}$ onwards).

As an example, the case of third-harmonic generation (THG) is considered [2–4]. The THG process can be described as $P_i^{3\omega} = \varepsilon_0 \chi_{ijkl}^{(3)} E_j^\omega E_k^\omega E_l^\omega$. $\chi_{ijkl}^{(3)}$ can be reduced to a simplified $\chi_{im}^{(3)}$ if Kleinman symmetry is applicable, with simplified matrix forms defined for various point groups analogous to the forms of the SHG tensor [2]. Once the appropriate $P_i^{3\omega}$ terms are generated by evaluating the THG equation using the $\vec{E}_{(iii)}$ fields as sources, $\vec{E}_{(iv)}$ can be obtained and one can continue with the rest of the framework as outlined in Section 3.

## Derivation of Radiation from Nonlinear Dipoles Using Green Dyadic Approach

The vector wave equation for the field radiated by a source is given by:

$$\nabla \times \nabla \times \vec{E}(\vec{r}) - k^2 \vec{E}(\vec{r}) = \vec{S}(\vec{r}) \qquad (S1)$$

where $\vec{S}(\vec{r})$ stands for the source.

The dyadic Green's function is given by:

$$\bar{\bar{G}}(\vec{r},\vec{r}') = \left(1 + \frac{1}{k^2}\nabla\nabla\cdot\right)g(\vec{r},\vec{r}')\bar{\bar{I}} \tag{S2}$$

$$\vec{E}(\vec{r}) = \int \bar{\bar{G}}(\vec{r},\vec{r}')\, \vec{S}(\vec{r}')d^3\vec{r}' = \int \left(1 + \frac{1}{k^2}\nabla\nabla\cdot\right)g(\vec{r},\vec{r}')\vec{S}(\vec{r}')d^3\vec{r}' \tag{S3}$$

In far field ($|\vec{r}-\vec{r}'| \to \infty$), the local wavefront at $\vec{r}$ due to a point source at $\vec{r}'$ can be approximated as a plane wave. Therefore $\nabla \sim i\vec{k}$, where $\vec{k} = k\hat{\kappa}$. Here $\hat{\kappa}$ is the unit vector representing the direction from the source point ($\vec{r}'$) to the field point ($\vec{r}$), or $\hat{\kappa} = \frac{\vec{r}-\vec{r}'}{|\vec{r}-\vec{r}'|}$.

$$\vec{E}(\vec{r}) \approx \int \left(1 + \frac{1}{k^2}ik\hat{\kappa}ik\hat{\kappa}\cdot\right)g(\vec{r},\vec{r}')\vec{S}(\vec{r}')d^3\vec{r}' = \int (\hat{\kappa}\cdot\hat{\kappa} - \hat{\kappa}\hat{\kappa}\cdot)g(\vec{r},\vec{r}')\vec{S}(\vec{r}')d^3\vec{r}'$$

$$= -\int \hat{\kappa}\times\hat{\kappa}\times g(\vec{r},\vec{r}')\vec{S}(\vec{r}')d^3\vec{r}' = -\int g(\vec{r},\vec{r}')\hat{\kappa}\times\hat{\kappa}\times \vec{S}(\vec{r}')d^3\vec{r}'$$

$$= -\int \frac{\exp(ik|\vec{r}-\vec{r}'|)}{4\pi|\vec{r}-\vec{r}'|^3}(\vec{r}-\vec{r}')\times(\vec{r}-\vec{r}')\times \vec{S}(\vec{r}')d^3\vec{r}' \tag{S4}$$

For a nonlinear dipole acting as a source for SHG, the source term $\vec{S}(\vec{r}') = \vec{P}^{2\omega}(\vec{r}')$, and $k(2\omega) = 2k(\omega) = 2k$ should be used. Hence one gets an expression for the SHG fields as:

$$\vec{E}(\vec{r}) = \int \frac{\exp(2ik|\vec{r}-\vec{r}'|)}{4\pi|\vec{r}-\vec{r}'|^3}\{(\vec{r}-\vec{r}')\times[(\vec{r}-\vec{r}')\times \vec{P}^{2\omega}(\vec{r}')]\}\, d^3r' \tag{S5}$$

### Dependence of SHG Signal on Sample Z-Position

The approach developed in section 3 of the main text is intended to capture the SHG response generated from a single surface in reflection. Aside from greatly simplifying calculations of the SHG response, such a condition is believed to be optimal for measurement when high NA lenses are in use. Figure S2 depicts the magnitude of SHG signal as a function of z-position of a 0.75 microscope objective as measured from a 1 mm thick z-cut α-quartz sample with both faces parallel and polished to an optical finish as received from the manufacturer (MTI Corporation).

An SHG response is generally only observed when either the top or bottom surface of the sample is in focus. As the microscope objective is translated along the Z-direction such that the focus is brought into the bulk of the sample, the SHG response decays and is no longer observed until the opposite interface is reached, suggesting that this phenomenon does not have an origin in multi-reflection effects. Naturally, as the focus of a probe is brought above a sample surface, the fluence will fall, and thus the local electric field strength and efficiency of the SHG process which goes as the square of the field. However, the decay in signal as the probe is focused into the depth of the sample reveals the importance of possessing an interface for reflection. Even if the amount of generated SHG is expected to increase as the beam waist is placed in the interior of the sample (up to a coherence length for the interaction), the generated SHG will not be properly recollected for detection if the beam is allowed to diverge before subsequent reflections. Focusing on the sample surface is thus recommended on the practical basis of maximizing the resolution of surface features as well as the magnitude of SHG signal collected.

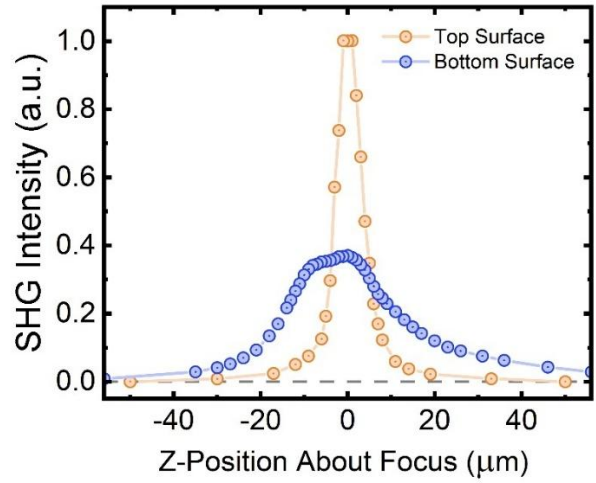

Fig. S2. SHG signal intensity versus microscope Z-position when the probe is initially focused on the top surface (orange) or bottom surface (blue) of a 1 mm thick z-cut α-quartz sample with both faces parallel and polished to an optical finish.

**Simulated and Experimental Polarimetry Curves for Material Case Studies**

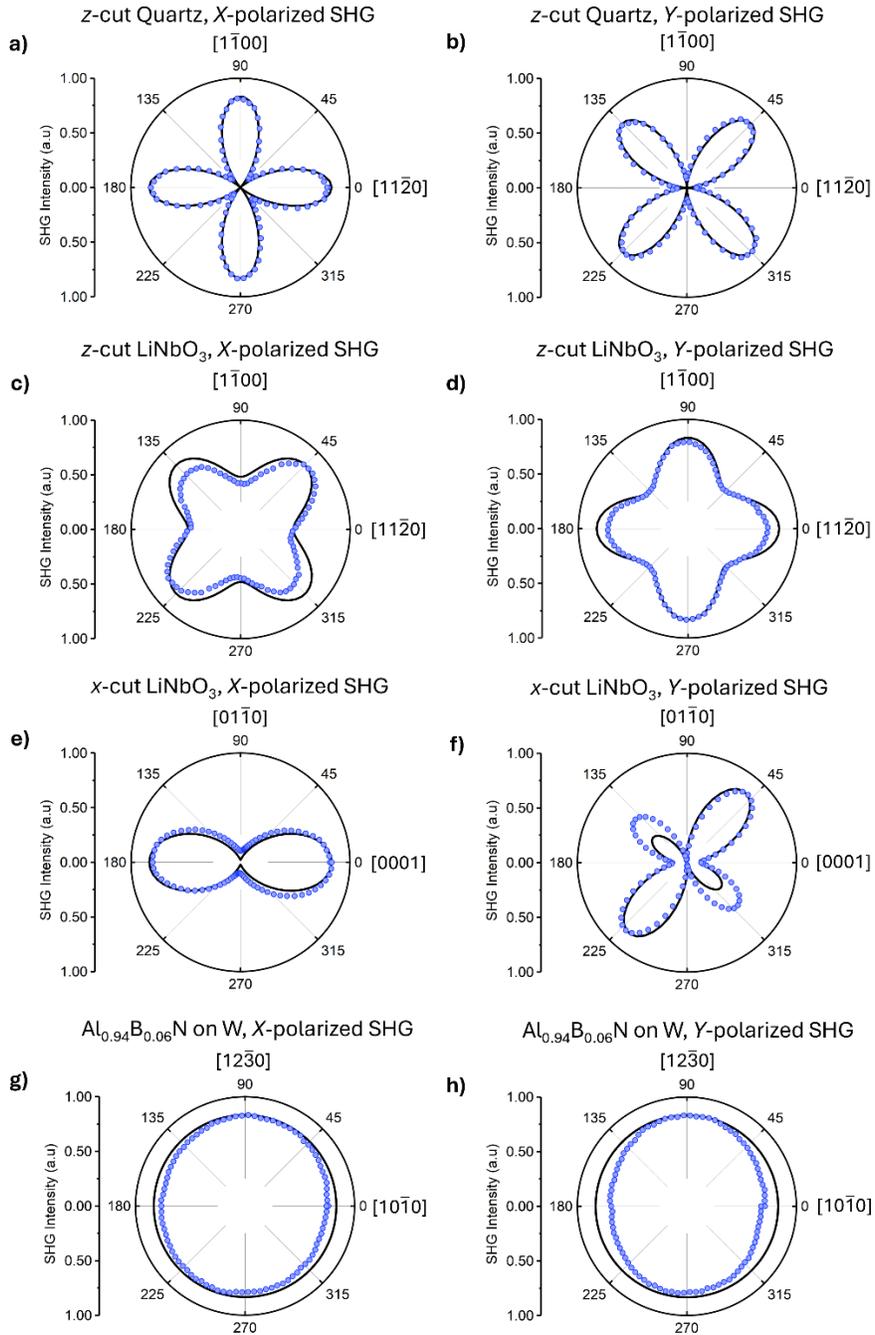

Fig. S3. Experimentally obtained (dotted, blue) and simulated (solid, black) SHG polarimetry curves obtained from single crystal *z*-cut quartz (a,b), *z*-cut LiNbO$_3$ (c,d), *x*-cut LiNbO$_3$, and *c*-axis oriented Al$_{0.94}$B$_{0.06}$N thin films on tungsten using a 0.75 NA objective. The *X*-direction of the setup is aligned with the [11$\bar{2}$0] crystal direction for $\alpha$-quartz, *z*-cut LiNbO$_3$ and Al$_{0.94}$B$_{0.06}$N, and the [0001] direction for *x*-cut LiNbO$_3$.

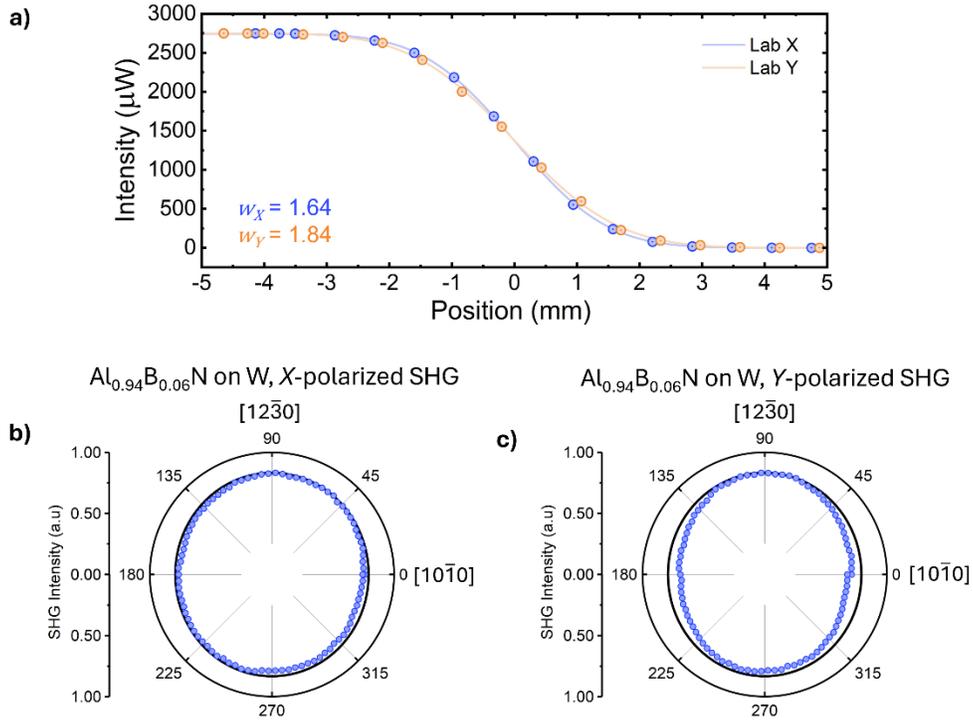

Fig. S4. a) Experimentally measured Gaussian beam profiles obtained using knife-edge measurements along the lab X (blue) and lab Y (orange) directions, fit to a form of, $I(r) = \frac{P}{2}[1 + \text{erf}(-r/w)]$. $X$-polarized (b) and $Y$-polarized (c) polarimetry curves from $c$-axis oriented $Al_{0.94}B_{0.06}N$ thin films on tungsten fit using a beam profile of the form $\vec{E} = \exp[(-x^2/2\sigma_x^2) - (y^2/2\sigma_y^2)]$, accounting for a slightly elliptical beam cross-section.

## Second Harmonic Signal Variation in Periodically Poled LiNbO₃ Domains

To investigate the origin of the different intensities in SHG signal observed from poled up and poled down domains in the periodically poled $LiNbO_3$ sample imaged in Fig. 5, a surface height map was obtained via atomic force microscope (AFM) measurements using a Bruker Dimension Icon operated in peak force tapping mode. The resultant AFM micrograph scan is illustrated in Fig. S4, showing height steps of roughly 340 nm. The beam waist and Rayleigh range of the probe beam used during SHG microscopy are calculated to be roughly 900nm and 3.2 μm respectively from knife edge measurements at multiple planes away from the focus. In calculating the expanded beam waist 340 nm from the focus using the standard equation for a Gaussian beam, the overall beam profile is found to expand by roughly 4% in area [5]. The observed intensity change in SHG signal from the two areas is roughly 5%, which is within reason given the calculated reduction in intensity for the larger spot size out of focus: $I_2/I_1 \approx S_2/S_1$ for identical electric fields where $I$ describes the beam intensity, $S$ the cross-sectional area, and 1 and 2 simply denote different planes.

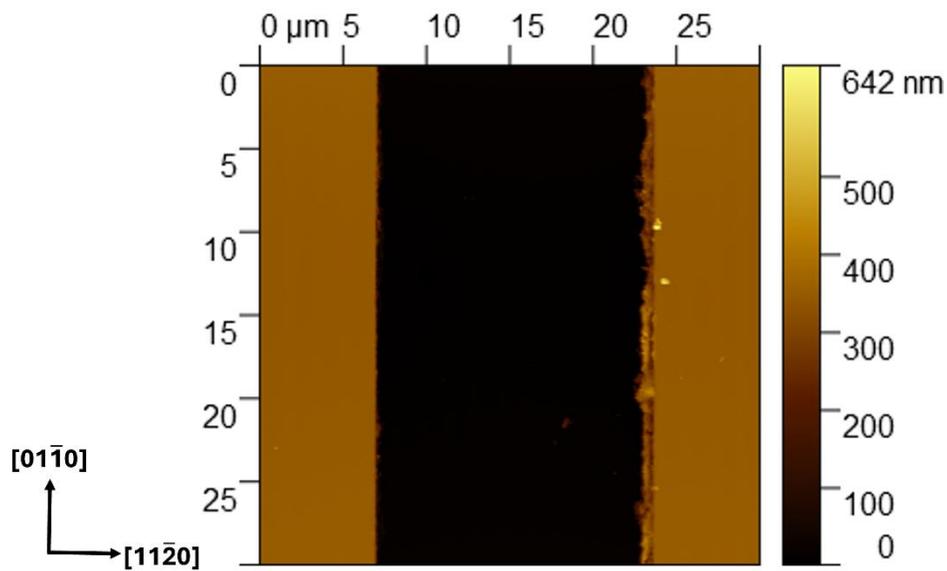

Fig. S5. AFM scan of a 30x30 μm surface region of the periodically poled LiNbO$_3$ sample imaged using SHG microscopy.

## SHG Polarimetry from Natural *a*1 and *a*2 Domains in BaTiO$_3$ Single Crystal

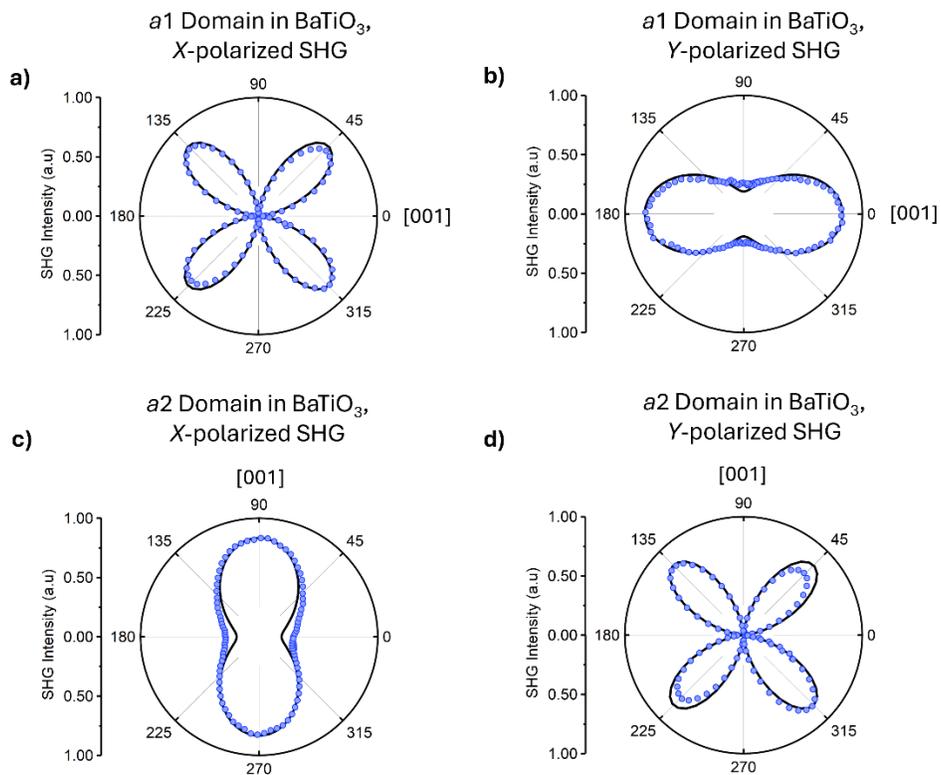

Fig. S6. Experimentally obtained (dotted, blue) and simulated (solid, black) SHG polarimetry curves obtained from natural *a*1 (a,b) and *a*2 (c,d) domains in a BaTiO$_3$ single crystal. *a*1 and *a*2 refer to the in-plane orientation of the polar axis as described in section 4.3 of the main text.

**Fitting Procedure for High NA Polarimetry Data**

Table S1. SHG Fitting Equations for Retrieving Ratios of *z*-cut LiNbO$_3$ Under 0.75 NA Lens

| Incident Polarization, Analyzer Orientation | Fitting Equation |
| --- | --- |
| 0° with respect to [11$\bar{2}$0], 0° with respect to [11$\bar{2}$0], | $I^{2\omega}_{0°,X} = 153.24\ d_{22}^2 + 4772.23\ d_{31}^2 + 4.73\ d_{33}^2 + 300.63\ d_{31}\ d_{33}$ |
| 45° with respect to [11$\bar{2}$0], 0° with respect to [11$\bar{2}$0], | $I^{2\omega}_{45°,X} = 27080.44\ d_{22}^2 + 4772.23\ d_{31}^2 + 4.73\ d_{33}^2 + 300.63\ d_{31}\ d_{33}$ |
| 0° with respect to [11$\bar{2}$0], 90° with respect to [11$\bar{2}$0], | $I^{2\omega}_{0°,Y} = 27080.44\ d_{22}^2 + 4772.23\ d_{31}^2 + 4.73\ d_{33}^2 + 300.63\ d_{31}\ d_{33}$ |
| 45° with respect to [11$\bar{2}$0], 90° with respect to [11$\bar{2}$0], | $I^{2\omega}_{45°,Y} = 153.24\ d_{22}^2 + 4772.23\ d_{31}^2 + 4.73\ d_{33}^2 + 300.63\ d_{31}\ d_{33}$ |

The fitting equations provided in Table S1 were used for quantifying SHG coefficient ratios shown in the main text Table 1. The incident polarization and analyzer orientations are described through their relative orientation with respect to the crystallographic [11$\bar{2}$0], which is aligned with the lab *X* direction as the setup is depicted in main text Fig. 1a. *C* is an amplitude scalar that is quantified by measurement against a reference sample where the tensor elements $d_{ij}$ are known.

The framework presented in section 3 is evaluated numerically, yielding solutions for polarization-dependent SHG intensities that depend on several factors, such as the input beam profile, the NA of the focusing objective, the SHG property tensor, and the material refractive index. The first two items listed depend on the experimental setup while the latter two are sample-dependent parameters that are often the subject of investigation. It is nontrivial to obtain an analytical solution parametrizing the effect of the experimental conditions chosen, so one must instead obtain a solution for the final SHG intensities that parametrizes the contributions from the SHG tensor elements for a fixed set of experimental conditions. The prescription for generating fitting equations for the SHG property tensor assuming that the experimental parameters and the sample refractive index are known beforehand proceeds as follows for the high NA case.

(1) Define an input beam profile and focusing lens NA that reflect the experimental conditions. The final fitting equations that follow will be valid only for these conditions. For example, one might use a simple Gaussian distribution for the probe with a spatial waist characterized using a knife-edge technique and a 0.75 NA focusing objective: $\vec{E} = \exp[-r^2/2\sigma^2]\ \hat{x}$

(2) Identify critical points, such as minima or maxima, for which parameterized solutions for the SHG response will be obtained. A simple plane-wave approximation-based polarimetry curve like those generated in section 4.1 of the main text, can aid in establishing initial expectations. An example of critical points selected for *z*-cut LiNbO$_3$ are shown in Fig. S6.

Several points on the polarimetry curve may be equivalent based on the symmetry of the polarimetry curve.

(3) Evaluate and inspect the SHG equation, $P_i^{2\omega} = \varepsilon_0 d_{ijk} E_j^\omega E_k^\omega$, using the appropriate form of the property tensor. If any $P_i^{2\omega}$ carries contribution from more than one tensor element, then the final SHG intensity will depend on cross terms carrying those coefficients as well as terms carrying only a single coefficient, originating from the squaring of fields to yield intensities. For example, for z-cut LiNbO$_3$ probed along 3 (3 || Z), $P_Z^{2\omega}$ is proportional to $d_{31}E_X^2 + d_{31}E_Y^2 + d_{33}E_Z^2$, and thus $I^{2\omega} \propto (P_Z^{2\omega})^2 = d_{31}{}^2 E_X^4 + d_{31}{}^2 E_Y^4 + d_{33}{}^2 E_Z^4 + 2d_{31}{}^2 E_X^2 E_Y^2 + 2d_{31}d_{33}E_X^2 E_Z^2 + 2d_{31}d_{33}E_Y^2 E_Z^2$. For a high NA lens, the final intensity $I^{2\omega}$ will depend on the sum of all $P_i^{2\omega}$ components and thus may be viewed as a parametrized sum of all such contributions: $I^{2\omega} = \sum \left( C_{d_{ij}} d_{ij}{}^2 + C_{d_{ij}d_{kl}} d_{ij}d_{kl} \right)$. Identifying the parametrized scaling parameters, $C$, for all terms is the objective of generating the fitting equations.

(4) Begin determining scaling parameters for terms with only a single coefficient, $C_{d_{ij}}$. Set all SHG coefficients save for one to be equal to 0. For example, z-cut LiNbO$_3$ possesses $d_{22}$, $d_{31}$, $d_{33}$, and $d_{15}$ as independent tensor coefficients. We may begin by setting $d_{31}$, $d_{33}$, and $d_{15}$ to 0 and $d_{22}$ to a dummy numerical value ($d_{22}$ = 1 pm/V).

(5) Evaluate the SHG polarimetry curves using the framework of the main text, noting the final SHG intensities at all critical points for a given value of the nonzero tensor element. Continue varying the nonzero tensor element until a relationship between the SHG intensity at every critical point and the tensor element being parametrized can be established. This relationship should be quadratic, requiring evaluation of at least 3 dummy values of the nonzero coefficient to fit.

(6) Repeat (3) and (4) for all independent tensor elements so that all $C_{d_{ij}} d_{ij}{}^2$ are obtained.

(7) Parametrize the contribution from crossed terms, $C_{d_{ij}d_{kl}}$, by setting all tensor elements except for the two involved in a given cross term to 0. For example, for z-cut LiNbO$_3$ we may begin by setting $d_{31}$ and $d_{33}$ to dummy numerical values while $d_{22}$ and $d_{15}$ are set to 0. Determine the relationship between the final SHG intensities at critical points and the product of the two tensor elements in the crossed terms similar to step (4). This relationship should be linear.

(8) Repeat (7) for all crossed terms identified in (6) so that all $C_{d_{ij}d_{kl}} d_{ij}d_{kl}$ are obtained.

(9) The final SHG intensities can be expressed as a sum of the parametrized contributions from all individual tensor elements and all crossed terms. For example, for z-cut LiNbO$_3$, the minima of the X-polarized polarimetry curves carries a fitting equation of the form $I_{0°,X}^{2\omega} = 153.24\ d_{22}{}^2 + 4772.23\ d_{31}{}^2 + 4.73\ d_{33}{}^2 + 300.63\ d_{31}\ d_{33}$, where $C_{d_{22}} = 153.24$ and $C_{d_{31}d_{33}} = 300.63$. Further equations are shown in Table S1. Terms in $d_{15}$ are absent due to the symmetry of the beam profile, as is discussed in detail regarding the analysis of SnP$_2$Se$_6$ in the main text. The leading constants $C$ are unique to the experimental conditions set in step (1). Note that this procedure is not sufficient for determining absolute SHG coefficients, only the ratios between fitted coefficients.

(10) The full set of SHG intensity equations serve as a set of fitting equations. Symmetry-equivalent experimental data points can then be fit against the full set of fitting equations in tandem using standard fitting procedures, like a root-mean-squared based optimization method.

The number of critical points chosen should be greater than or equal to the number of independent tensor elements being fit.

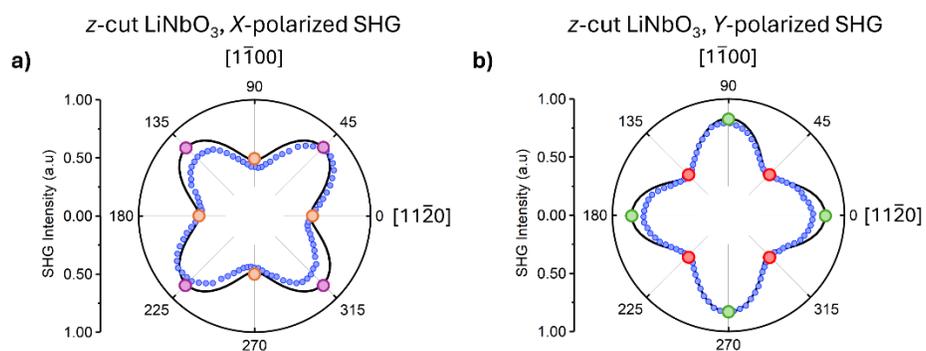

Fig. S7. Critical points for fitting identified in *z*-cut LiNbO$_3$ polarimetry curves, for an analyzer oriented along the lab X direction (a) and lab Y direction (b). Parametrized SHG intensity equations for these points are given in Table S1.

## SHG Microscope Images from an Additional Region of SnP$_2$Se$_6$ Single Crystal

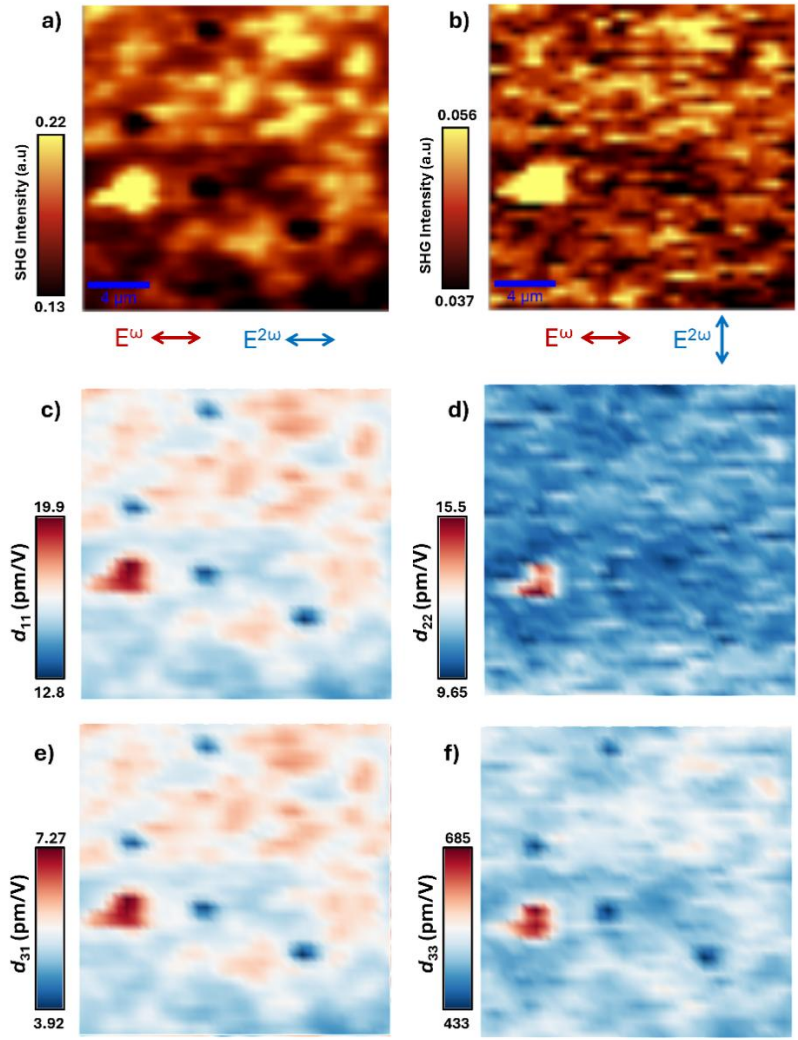

Fig. S8. Experimentally obtained SHG intensity maps of $SnP_2Se_6$ obtained at two critical configurations for coefficient ratio extraction: lab *X* fundamental polarization in with a) lab *X* or b) lab *Y* SHG polarization detected. The *X*-direction of the setup is aligned with the $[2\bar{1}\bar{1}0]$ crystal direction and the *Y*-direction aligned with $[01\bar{1}0]$. c-f) SHG coefficient maps obtained by collecting maps at additional critical points and performing simultaneous point-by-point fits to obtain the location dependence of property tensor coefficient ratios. The average value of the $d_{11}$ coefficient map was set to the value obtained via 0.12 NA measurements and used to calibrate the values of other coefficients.

## References

1. M. Leutenegger, R. Rao, R. A. Leitgeb, and T. Lasser, "Fast focus field calculations," Opt. Express **14**(23), 11277–11291 (2006).
2. X. Yang and S. Xie, "Expression of third-order effective nonlinear susceptibility for third-harmonic generation in crystals," Appl Opt **34**(27), 6130–6135 (1995).
3. G. H. C. New and J. F. Ward, "Optical Third-Harmonic Generation in Gases," Phys Rev Lett **19**(10), 556–559 (1967).
4. Robert W. Boyd, *Nonlinear Optics* (Elsevier, 2003).
5. A. Yariv and P. Yeh, *Optical Waves in Crystals: Propagation and Control of Laser Radiation* (2002).